# Data-driven modeling and supervisory control system optimization for plug-in hybrid electric vehicles

Hao Zhang [a], Nuo Lei [a], Boli Chen [b], Bingbing Li [c], Rulong Li [d], Zhi Wang [a,*]

[a] State Key Laboratory of Intelligent Green Vehicle and Mobility, Tsinghua University, Beijing 100084, China

[b] Department of Electronic and Electrical Engineering, University College London, London WC1E 6BT, UK

[c] Department of Mechanical Engineering, Southeast University, Nanjing 211189, China

[d] Dongfeng Motor Corporation Ltd., Wuhan 430058, China

*Corresponding author: Zhi Wang

Professor

State Key Laboratory of Intelligent Green Vehicle and Mobility

School of Vehicle and Mobility, Tsinghua University

Beijing, 100084, China

Email: wangzhi@tsinghua.edu.cn

## Highlights

- Enhancing RL applicability through data-driven modeling and reliable control framework.
- Data-driven model error calibration toolchain for high-fidelity environment modeling.
- Learning-based energy management combined with real-time optimization method.
- Corrects flawed short-term evaluation of equivalent factors using horizon-extended RL.
- The proposed framework enhances disturbance resistance without damaging optimality.

## Abstract

Learning-based intelligent energy management systems for plug-in hybrid electric vehicles (PHEVs) are crucial for achieving efficient energy utilization. However, their application faces system reliability challenges in the real world, which prevents widespread acceptance by original equipment manufacturers (OEMs). This paper begins by establishing a PHEV model based on physical and data-driven models, focusing on the high-fidelity training environment. It then proposes a real-vehicle application-oriented control framework, combining horizon-extended reinforcement learning (RL)-based energy management with the equivalent consumption minimization strategy (ECMS) to enhance practical applicability, and improves the flawed method of equivalent factor evaluation based on instantaneous driving cycle and powertrain states found in existing research. Finally, comprehensive simulation and hardware-in-the-loop validation are carried out which demonstrates the advantages of the proposed control framework in fuel economy over adaptive-ECMS and rule-based strategies. Compared to conventional RL architectures that directly control powertrain components, the proposed control method not only achieves similar optimality but also significantly enhances the disturbance resistance of the energy management system, providing an effective control framework for RL-based energy management strategies aimed at real-vehicle applications by OEMs.

## 1. Introduction

Energy conservation and de-carbonization are essential to the sustainable development of global industry and economy [1]. China has announced its aim to reach the peak of carbon emission by 2030 and strives to achieve carbon neutrality by 2060 [2]. The automotive industry has always been a key focus in carbon emission management, and effectively solving vehicle energy consumption and emission issues is an important task for the industry [3, 4]. Promoting the development of energy-saving and new energy vehicles has become one of the core strategies [5]. Plug-in hybrid electric vehicles (PHEVs) represent a mature and realistic energy-saving technology for the foreseeable future [6]. Generally, original equipment manufacturers (OEMs) have similar development capabilities in PHEV hybrid configuration and hardware [7], possessing relatively mature powertrain platforms and vehicle products [8], but further research and development are still needed in intelligent energy management system (EMS) [9].

The widely used energy management systems in industry are rule-based (RB) and optimization-based methods, and the latter one can be divided into global optimization and instantaneous optimization [10]. The global optimization strategies solve the optimal power distribution of hybrid vehicles under predetermined conditions using optimal control methods [11]. In contrast, the principle of instantaneous optimization-based energy management is to transform the optimization in infinite time domain into an instantaneous or finite time domain optimization problem [12]. The prerequisite for global optimization of energy management is the known power demand of the entire driving cycle [13]. Addressing the real-time control requirements, scholars have proposed energy management strategies based on instantaneous optimization, without requiring prior knowledge of the complete cycle conditions [14], thus not limited by specific cycle conditions and featuring smaller computational requirements and feasibility for real-time applications, among which the equivalent consumption minimization strategy (ECMS) is one of the most famous methods [15].

As an online implementation of the Pontryagin minimum principle (PMP), the ECMS control strategy has been validated on parallel hybrid vehicle products, proving its excellent optimization performance and real-time applicability [16]. ECMS optimally distributes the current demand power across multiple power sources based on the conversion coefficient of fuel consumption and electricity consumption [17]. This conversion coefficient, commonly known as the equivalent factor (EF), serves the same function as the co-state variable in PMP [18]. The research focus of ECMS is on how to accurately estimate the EF, with its estimation methods summarized into two categories: 1) using global optimization algorithms' offline solutions to determine the optimal EF, set as a constant throughout the driving cycle [19]; 2) using online estimation methods to update the EF in real-time . For offline EF estimation methods, if the complete information of the cycle conditions is known [20], the best constant EF value can be found, solvable by global optimization algorithms such as genetic algorithms, dynamic programming (DP), and the shooting algorithm [21]. Although driving cycle characteristics can be estimated through prediction and pattern recognition methods, a constant EF is hard to adapt to complex driving cycles [22]. A common solution to this problem is feedback control of EF based on the error between the actual value of the battery state of charge (SOC) and the SOC reference value, including proportional control, proportional integral (PI) control and non-linear control methods, to maintain SOC at a specified level [23]. For instance, Musardo et al. proposed the adaptive-equivalent consumption minimization strategy (A-ECMS), adjusting the equivalent factor in real-time according to different road conditions to further approach the global optimum [24]. Yang et al. combined driver style recognition with the A-ECMS algorithm for energy management in hybrid buses, reducing fuel consumption by 5.29%. Although A-ECMS is a more feasible real-time optimization control strategy, it is essentially a local optimization thus cannot guarantee global optimality [25].

Given the complexity of the design of real-time optimization control methods, the substantial computing resources and their poor adaptability to actual road conditions, the academic and industrial communities are gradually focusing on learning-based EMS [26]. Benefiting from the powerful high-dimensional fitting capabilities of neural networks, learning-based energy management strategies can be divided into supervised learning, unsupervised learning, and reinforcement learning (RL) [27]. Supervised learning requires a large amount of global optimal state-action trajectories solved by offline optimization algorithms like DP and PMP as sample data [28]. The defect of supervised learning control systems is the difficulty in obtaining a large amount of offline optimal sample data, and the complexity and constant change of actual driving conditions [29]. Unsupervised learning relies on the distribution and structure of the data itself, however, it's challenging to define a quantifiable measurement to assess the quality of PHEV energy management [30]. Therefore, it leads to potential inefficiency in adapting EMS to varying driving cycles or even failing to learn effective control policy [31]. Compared to supervised and unsupervised learning, RL, through the interaction between agents and environments and adjustment of strategies through rewards and punishments, has advantages in handling the sequential decision problem of PHEV energy management [32]. It can better adapt to unknown and changing driving conditions, improving fuel efficiency and adaptability [33].

The basic principle of RL also relies on the iterative solution of the Bellman equation, meaning the problem studied needs to conform to the Markov decision process (MDP) [34]. The optimal energy management strategy is defined as the control trajectory that maximizes long-term returns rather than focusing on rewards at each moment [35], which is a core difference from traditional instantaneous optimization control strategies, with some related studies listed in Table 1. Liu et al. pioneered in proposing an energy management system for series hybrid vehicles based on Q-learning and compared the differences in energy management applications between model-free Q-learning algorithms and model-based Dyna-Q learning algorithms [36]. Lee et al. compared EMSs based on dynamic programming and RL, proving RL could achieve optimal solutions approximating those of DP in infinite time-domain control [37]. Furthermore, to address the problem of high-dimensional state inputs, scholars have combined deep neural networks with RL, proposing deep reinforcement learning (DRL) algorithms. Wu et al. adopted deep Q-Learning (DQL)-based energy management strategy for power-split hybrid buses, showing DQL outperforms traditional Q-learning algorithms [38]. Shuai et al. applied the double Q-learning algorithm for hybrid vehicle energy management, improving RL's stability and convergence speed by avoiding overestimation, achieving a 7.1% fuel-saving rate compared to conventional Q-learning [39]. Although the effectiveness of RL in energy management tasks is widely recognized, most RL-based energy management strategies are concentrated on simulation studies [40]. The effectiveness of these algorithms under actual vehicle conditions has not been validated, and their control effects depend on the consistency between the algorithm training environment and the real environment in terms of vehicle dynamic characteristics [41]. The application of EMSs from simulation environments to real environments is known as strategy transfer. If there is a significant discrepancy between the model and the real environment, it could lead to the failure of strategy transfer, greatly reducing the control effect of RL strategies [42, 43].

**Table 1**
Research progress on reinforcement learning-based energy management

| Research Team | Configuration | Algorithm | Agent action |
|---|---|---|---|
| Beijing Institute of Technology [44] | Power-split | DDPG | Engine power |
| Tsinghua University [45] | Power-split | DQL | Engine power |
| Chongqing University [35] | Power-split | Q-Learning | Engine power |
| Southeast University [46] | Series | PPO | Engine power |
| University of Birmingham [39] | Series | Double Q-Learning | Engine power |
| University of Oklahoma [47] | Parallel | Q-Learning | Power distribution ratio |
| Sophia University [48] | Parallel | PPO | Drive mode, engine torque |
| Seoul National University [37] | Parallel | Q-Learning | Engine torque |

It is evident that although RL-based energy management strategies have been extensively studied, the industry still predominantly uses rule-based EMSs [49]. Only a few OEMs have attempted to mass-produce instantaneous optimization-based energy management systems, while learning-based energy management control methods are rarely applied [50]. This indicates that intelligent EMSs have not yet achieved complete breakthrough, and there is a significant gap between scientific research and engineering applications [51]. Currently, there are mainly two problems: 1) The high-fidelity vehicle models have not been applied to RL agent training that consider system dynamic responses. Current research often uses overly simplified models or introduces too many assumptions, making it difficult to accurately simulate the complex dynamic behavior of key components [52]. 2) The absence of a reliable, vehicle-applicable control framework for RL-based energy management. In existing RL-based energy management methods, the agent's output actions directly control powertrain components such as transmissions and engines [53]. However, the control decisions generated by neural networks lack interpretability and reliability, urgently requiring the development of a RL-based energy management control framework with reliance on PHEV digital twin modeling [54].

To the authors' knowledge, current research on EMS has not yet adopted high-fidelity environment models using data-driven approaches, and there is a lack of reliable RL energy management control framework. Moreover, only a few studies have explored combining RL with ECMS, but the state selection of instantaneous driving cycles and powertrain state inputs in RL agent does not enable correct evaluation of ECMS equivalent factors. In response to these problems, this paper proposes a physical model and data hybrid-driven calibration method for effective correction of model errors, tailored to the high-fidelity training environment required by RL, as shown in Fig. 1. By utilizing real-vehicle chassis test data and machine learning methods, a high-accuracy model of the hybrid powertrain is established. Subsequently, an energy management method based on the ECMS and RL is proposed with clear physical meaning and strong disturbance resistance, realizing real-time estimation of the optimal equivalent factor. Finally, the methods are validated based on simulation and hardware-in-the-loop (HIL) tests, validating the control system's disturbance resistance, providing a reliable framework for the engineering integration of RL.

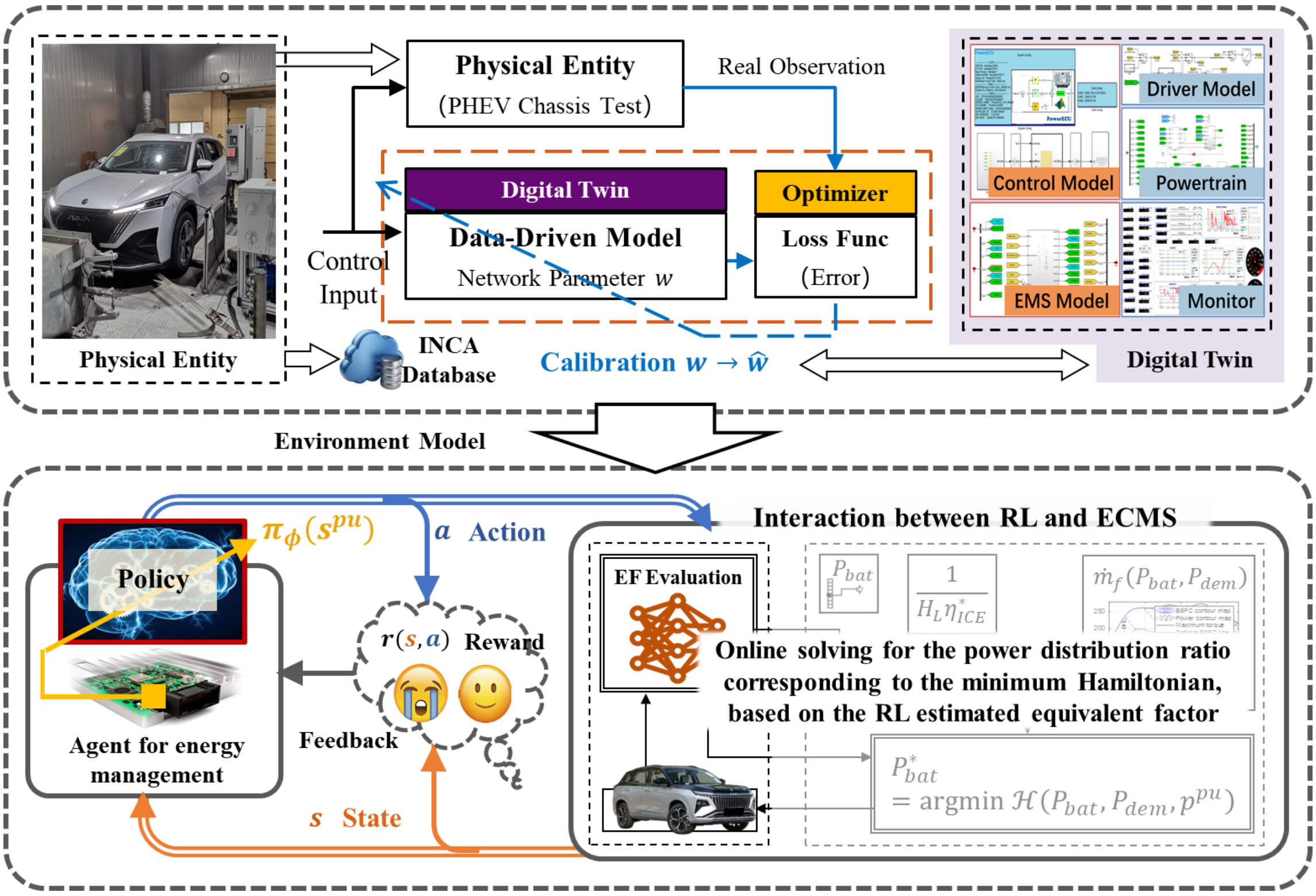

Fig. 1. The overall investigation procedure of the dedicated modeling and control framework for RL-based energy management of PHEVs.

The remainder of this paper is structured as follows: Section 2 introduces the real-vehicle experimental test and physical modeling of the vehicle dynamic model. Section 3 describes the data-driven model error calibration method and the horizon-extended RL merged with real-time optimization framework. Section 4 presents the HIL test setup and benchmark EMSs. Section 5 discusses the modeling results, learning ability, control performance and disturbance resistance capability. Section 6 summarizes the key findings.

## 2. Powertrain physical model and experimental calibration

### 2.1. PHEV powertrain configuration and modeling

The PHEV studied in this article was developed by Dongfeng Motor Corporation, which utilizes a planetary gear set and two shifting drums. The overall structure of the powertrain is shown in Fig. 2, where the planetary gear set serves as the power distribution device between the engine and the generator. The physical model of the PHEV powertrain is explained as follows:

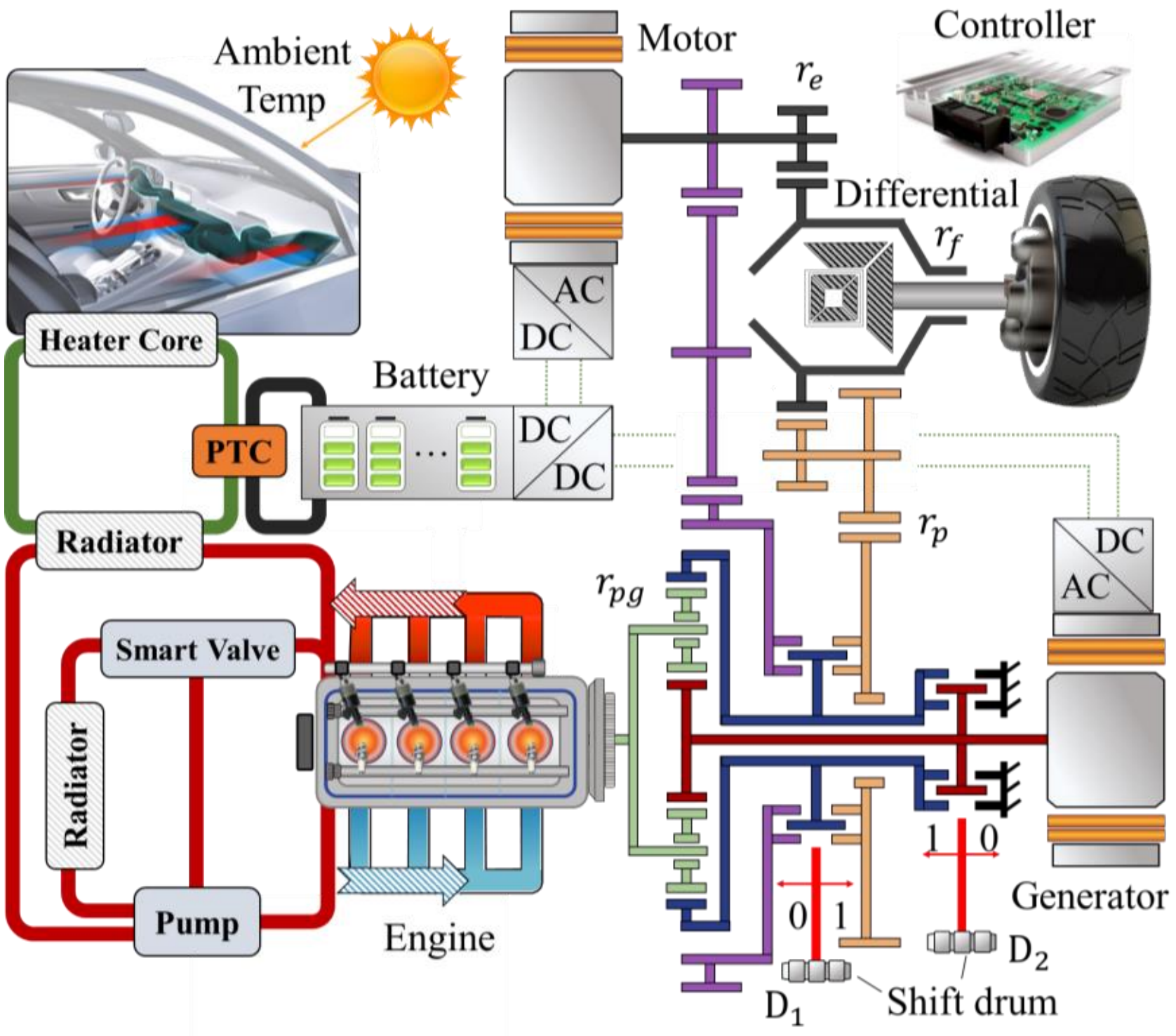

Fig. 2. Topology of the plug-in hybrid electric powertrain.

$$\begin{cases} r_f r_e T_{mot} = T_{dem} \\ \omega_{ICE} = \omega_{gen} = 0 \end{cases}, \text{if} \begin{bmatrix} S_1 \\ S_2 \end{bmatrix} = \begin{bmatrix} 0 \\ 0 \end{bmatrix}$$
$$\begin{cases} r_f r_e T_{mot} = T_{dem} \\ \omega_{ICE} = \omega_{gen} \end{cases}, \text{if} \begin{bmatrix} S_1 \\ S_2 \end{bmatrix} = \begin{bmatrix} 0 \\ 1 \end{bmatrix}$$
$$\begin{cases} r_f r_p T_{ICE} + r_f r_e T_{mot} = T_{dem} \\ (1 + r_{pg}) r_e \omega_{ICE} = r_p r_{pg} \omega_{mot} \end{cases}, \text{if} \begin{bmatrix} S_1 \\ S_2 \end{bmatrix} = \begin{bmatrix} 1 \\ 1 \end{bmatrix} \quad (1)$$
$$\begin{cases} \dfrac{r_f r_p r_{pg}}{(1 + r_p r_{pg})} T_{ICE} + r_f r_e T_{mot} = T_{dem} \\ (1 + r_{pg}) \omega_{ICE} = \dfrac{r_p r_{pg}}{r_e} \omega_{mot} + \omega_{gen} \end{cases}, \text{if} \begin{bmatrix} S_1 \\ S_2 \end{bmatrix} = \begin{bmatrix} 1 \\ 0 \end{bmatrix}$$

where $T_{dem}$ is defined as the torque demand at the wheel, and $r$ means the radius of the wheel. Also, $\omega_{ICE}$, $\omega_{gen}$, and $\omega_{mot}$ are the angular speeds of the engine, generator, and electric motor, respectively, while $T_{ICE}$, $T_{gen}$, and $T_{mot}$ represent the torques of the components. Moreover, $r_{pg}$ and $r_f$ represent the transmission ratios of the planetary gears and the main reducer. The detailed definition and parameters can be referred to the authors' previous work.

### 2.2. *Vehicle chassis test and physical modeling of multi-mode hybrid powertrain*

The vehicle chassis test bench shown in Fig. 3 is used to acquire experimental data of the PHEV for high-fidelity vehicle modeling. This test bench consists of the tested vehicle, dynamometer, cooling fan, and monitoring system. The dynamometer is used to realize the driving load and the operation data in vehicle control unit (VCU) and engine control unit (ECU) are recorded and monitored through the INCA software. Specific parameters of the PHEV are listed in Table 1.

**Table 1**
Vehicle technical specifications

| Component | Parameters | Value |
|---|---|---|
| Vehicle | Total Mass | 1830 kg |
| | Rolling Resistance | 0.013 |
| | Air Resistance Coefficient | 0.325 |
| | Frontal Area | 2.3 $m^2$ |
| IC Engine | Maximum Power | 120 kW |
| | Maximum Torque | 270 N·m |
| | Maximum Speed | 5200 rpm |
| Generator (MG1) | Maximum Power | 50 kW |
| | Maximum Torque | 115 N·m |
| | Maximum Speed | 9000 rpm |
| Motor (MG2) | Maximum Power | 70 kW |
| | Maximum Torque | 150 N·m |
| | Maximum Speed | 6000 rpm |
| Battery | Battery Capacity | 85 A·h |
| | Nominal Voltage | 345 V |
| | Number of cells in series | 105 |

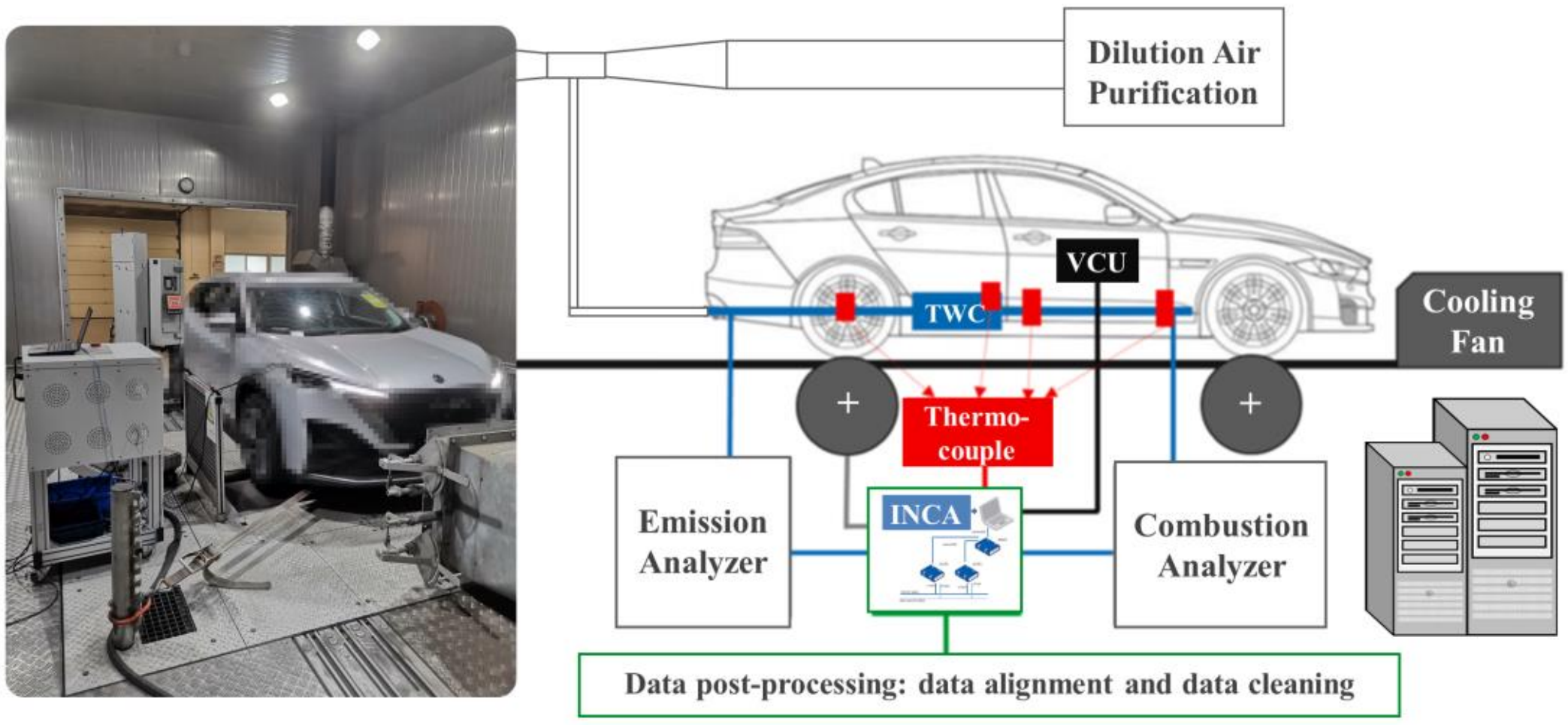

Fig. 3. The vehicle chassis dynamometer test bench.

The map-based quasi-static engine model is shown in Fig. 4, obtained through the vehicle chassis tests and characterizes the engine's working characteristics. Furthermore, a dynamic model of the engine is established based on the principle of the mean value model. The transient engine model is expressed from Eq. (2) to Eq. (10), which considers the torque command $T_{\mathrm{ICE}}^{\mathrm{cmd}}$, throttle position $\theta$, spark advance angle $SA$, fuel injection timing $C_{\mathrm{f}}$, EGR valve opening $EGR_{\mathrm{vp}}$, air-fuel ratio $\lambda$, coolant temperature $T_{\mathrm{ICE}}^{\mathrm{cd}}$, intake/exhaust temperatures $T_{\mathrm{int}}/T_{\mathrm{exh}}$ and intake/exhaust pressures $p_{\mathrm{int}}/p_{\mathrm{exh}}$.

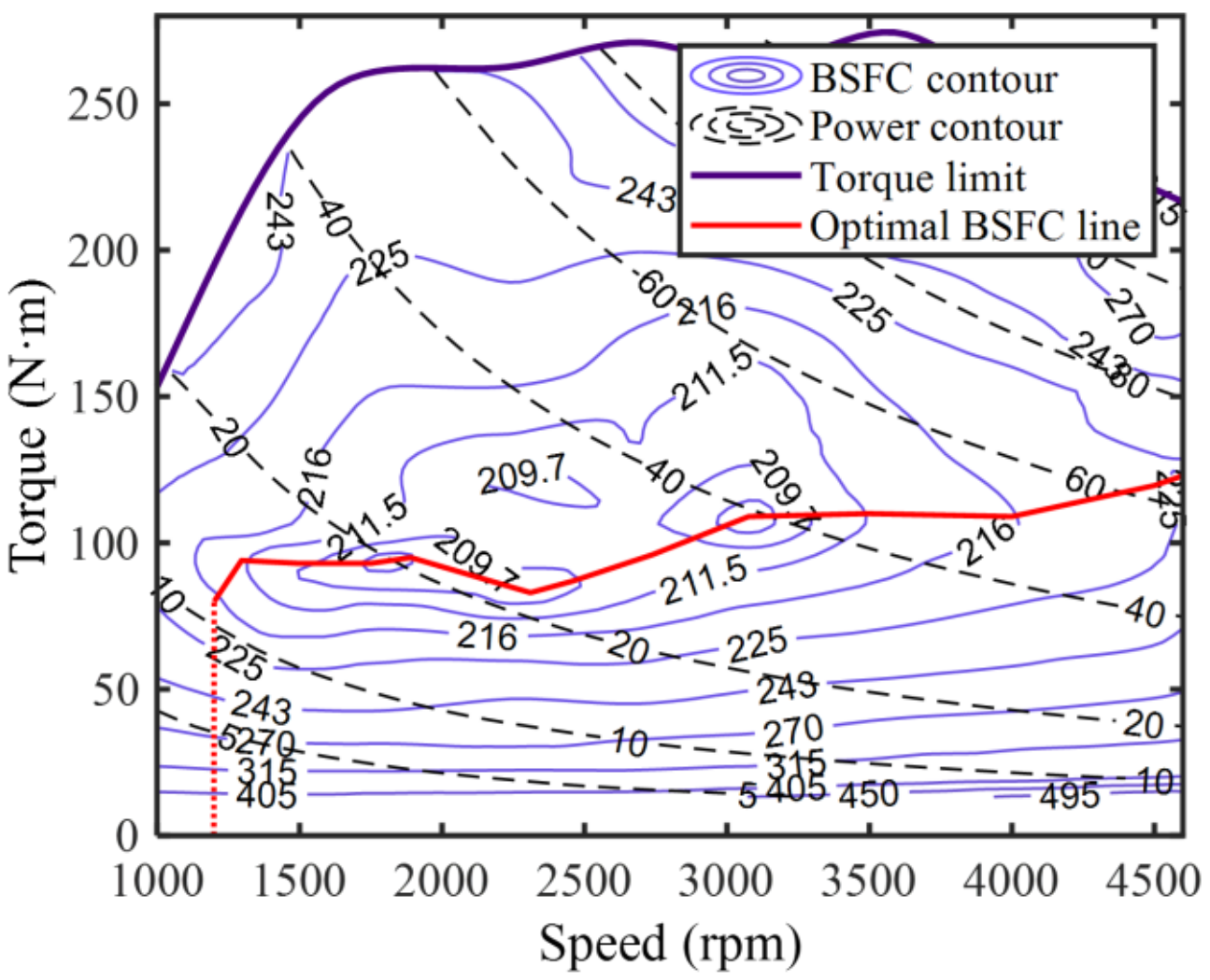

Fig. 4. Engine brake specific fuel consumption map.

$$\dot{m}_{\mathrm{fuel}}^{\mathrm{d}} = \frac{\dot{m}_{\mathrm{at}}}{\lambda L_{\mathrm{th}}}\left[1 + \mu\left(\frac{T_{\mathrm{ICE}}^{\mathrm{c*}} - T_{\mathrm{ICE}}^{\mathrm{cd}}}{T_{\mathrm{ICE}}^{\mathrm{c*}} - T_{\mathrm{ICE}}^{\mathrm{c0}}}\right)^{\xi}\right] \tag{2}$$

$$\begin{cases} \dot{m}_{\mathrm{at}} = f(\theta, p_{\mathrm{int}}) = \dfrac{C_{\mathrm{DV}}\gamma^{0.5}}{\sqrt{R_{\mathrm{m}}T_0}} A(\theta) p_{\mathrm{int}} \sqrt{\dfrac{2\gamma}{\gamma - 1}\left[1 - \left(\dfrac{p_{\mathrm{int}}}{p_0}\right)^{\frac{\gamma-1}{\gamma}}\right]} \\ A(\theta) = 1 - \cos(\theta - \theta_0) \end{cases} \tag{3}$$

$$\dot{m}_{\mathrm{EGR}} = f\left(EGR_{\mathrm{vp}}, p_{\mathrm{exh}}\right) \tag{4}$$

$$\dot{m}_{\mathrm{ac}} = \dot{m}_{\mathrm{at}} + \dot{m}_{\mathrm{EGR}} - \frac{p_{\mathrm{int}}V_{\mathrm{int}}}{R_{\mathrm{m}}T_{\mathrm{int}}} \tag{5}$$

$$\lambda = \frac{\dot{m}_{\mathrm{fuel}}^{\mathrm{d}}}{\dot{m}_{\mathrm{at}} + \dot{m}_{\mathrm{EGR}} - \dot{m}_{\mathrm{ac}}} \tag{6}$$

$$p_{\mathrm{int}} = \frac{c_1\dot{m}_{\mathrm{at}} + c_3\omega_{\mathrm{ICE}}}{c_2\omega_{\mathrm{ICE}}} \tag{7}$$

$$p_{\mathrm{exh}} = \epsilon p_{\mathrm{int}} \tag{8}$$

$$\eta_{\mathrm{vol}} = s_1 + s_2\omega_{\mathrm{ICE}} + s_3{\omega_{\mathrm{ICE}}}^3 + s_4 p_{\mathrm{int}} \tag{9}$$

$$T_{\mathrm{ICE}}^{\mathrm{cd}} = \frac{\left(\dot{m}_{\mathrm{f}}^{\mathrm{d}}\mathrm{LHV} - P_{\mathrm{ICE}}\right) - \left(\dot{Q}_{\mathrm{exh}} + \dot{Q}_{\mathrm{rad}} + \dot{Q}_{\mathrm{cab}}\right)}{m_{\mathrm{ICE}}C_{\mathrm{ICE}}} \tag{10}$$

where $\dot{m}_{\mathrm{at}}$, $\dot{m}_{\mathrm{EGR}}$ and $\dot{m}_{\mathrm{ac}}$ represent the airflow through the throttle, EGR rate, and cylinder air charge, respectively. $L_{\mathrm{th}}$ is the chemical equivalence air-fuel ratio, $C_{\mathrm{DV}}$ represents the dissipation coefficient at the valve, $\gamma$ and $R$ are the adiabatic coefficient and universal gas constant, respectively. $T_0$ and $p_0$ refer to the ambient temperature and pressure, $A(\theta)$ is the throttle area as a function of throttle position, $V_{\mathrm{int}}$ is the intake manifold volume, $\epsilon$ is the volumetric compression ratio, $VT$ indicates valve timing, and $\eta_{\mathrm{vol}}$ is the volumetric efficiency, LHV and $T_{\mathrm{ICE}}^{\mathrm{c*}}$ are the lower heating value of gasoline and the target coolant temperature, respectively. The heat carried away by exhaust, through air convection, and supplied to the cabin through the radiator are represented by $\dot{Q}_{\text{exh}}$, $\dot{Q}_{\text{rad}}$, and $\dot{Q}_{\text{cab}}$, respectively. $c_{1\sim3}$, $s_{1\sim4}$ are calibration parameters obtained by regression estimation of engine bench test data.

The hybrid powertrain studied in this paper includes a P1 generator connected to the engine and a P3 motor connected to the drive shaft. Bench tests of the electric machines provide their operating maps, see Fig. 4, and the efficiency can be obtained from these maps based on their speed $\omega_{\text{MG}}$ and torque $T_{\text{MG}}$. shows the efficiency map charts for both the P1 and P3 motors. The equivalent circuit model is adopted in this work for battery modeling, and the internal resistance and open circuit voltage are obtained from the real-vehicle tests, as shown in Fig. 5. The detailed mathematical expression of the electric machines and battery modeling can be found in the authors' previous work [52].

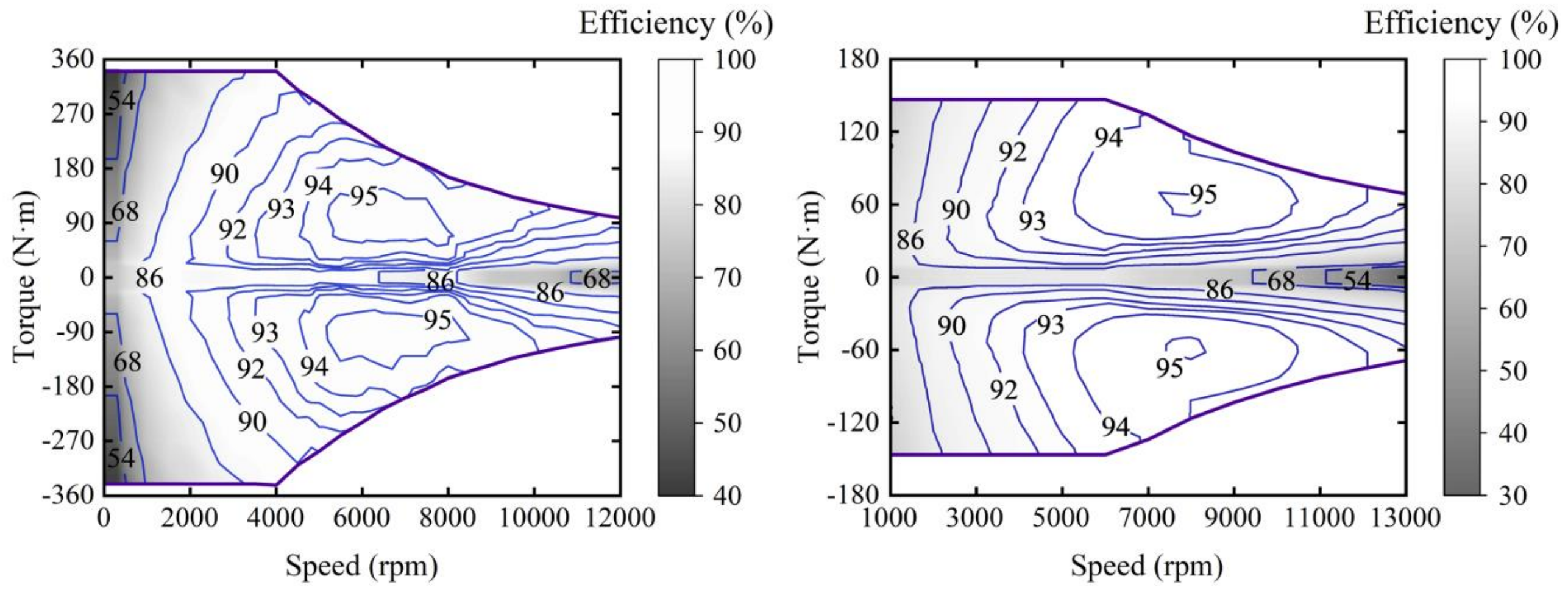

(a) (b)

Fig. 5. Electric machine efficiency maps of drive motor (a) and generator (b).

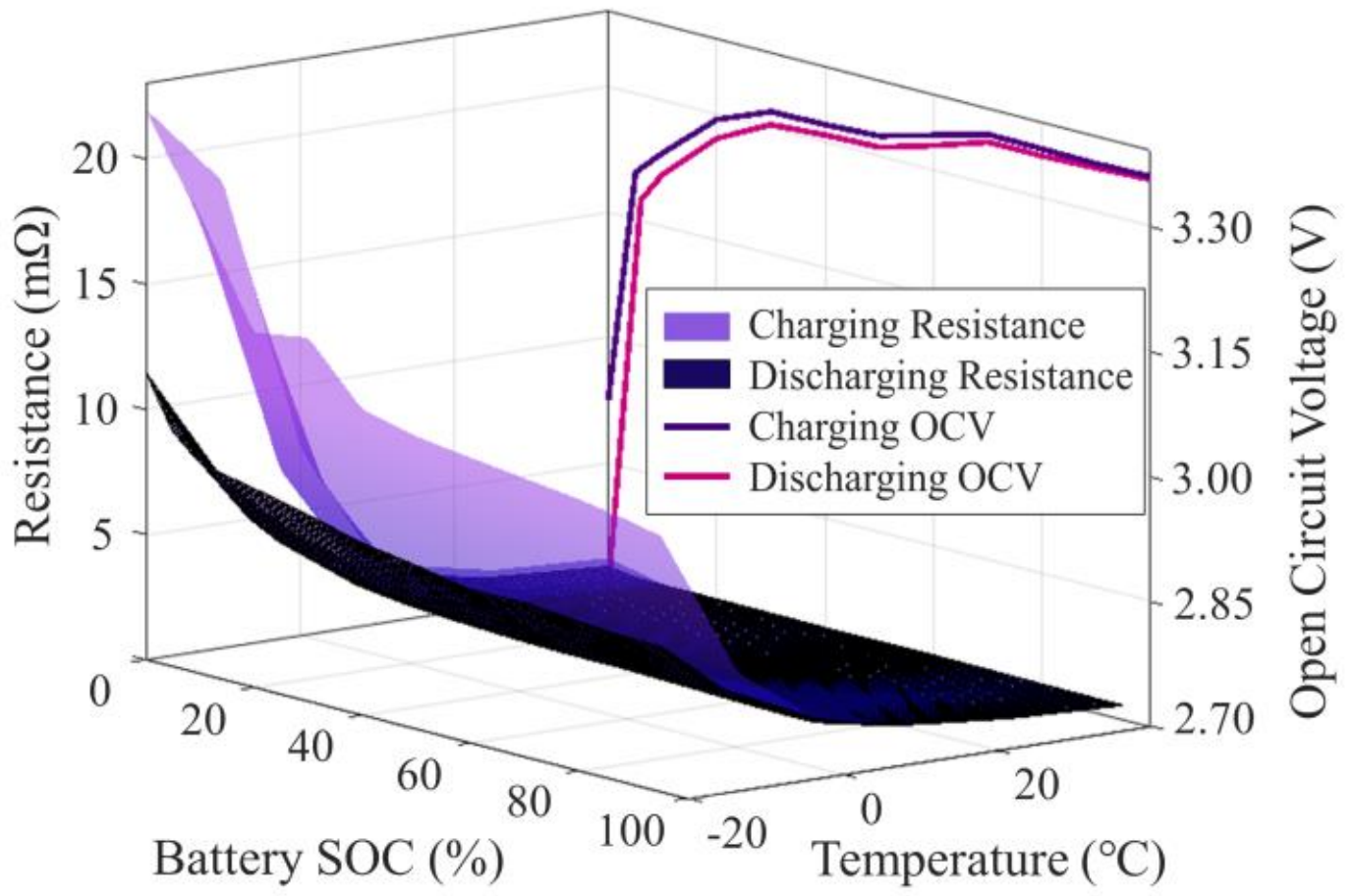

Fig. 6. Open circuit voltage and internal resistance of battery.

## 3. Dedicated modeling and control framework for RL-based EMS

### *3.1. Data-driven model error calibration toolchain for high-fidelity environment modeling*

The high-fidelity engine modeling is crucial to the environment establishment for the training of RL-based EMS, since the fuel consumption of PHEV originates from the internal combustion engine.

Based on the experimental maps and physical dynamic model presented in Section 2.2, the physical model and data hybrid-driven fuel consumption model is further proposed combining the quasi-steady-state map, physical dynamic models as well as the data-driven error correction model. In this work, the data-driven correction model is used to fit the deviation $\Delta\dot{m}_{\mathrm{fuel}}$ between the real data $\dot{m}_{\mathrm{fuel}}^{\mathrm{r}}$ and the physical model presented in Section 2.2, i.e., the averaged value of the map and physical dynamic model, denoted as $\dot{m}_{\mathrm{fuel}}^{\mathrm{a}}$. To explain this, Eq. (11) presents the relationship among $\Delta\dot{m}_{\mathrm{fuel}}$, $\dot{m}_{\mathrm{fuel}}^{\mathrm{r}}$ and $\dot{m}_{\mathrm{fuel}}^{\mathrm{a}}$. By providing the same control inputs to the current simulation model and utilizing the error between the test data and the averaged model output $\dot{m}_{\mathrm{fuel}}^{\mathrm{a}}$, the neural network can be tuned, without needing to modify the physical parameters of the dynamic model, as shown in Fig. 7.

$$\dot{m}_{\mathrm{fuel}} = \dot{m}_{\mathrm{fuel}}^{\mathrm{a}} + \Delta\dot{m}_{\mathrm{fuel}} = \frac{\dot{m}_{\mathrm{fuel}}^{\mathrm{d}} + \dot{m}_{\mathrm{fuel}}^{\mathrm{q}}}{2} + \Delta\dot{m}_{\mathrm{fuel}} \tag{11}$$

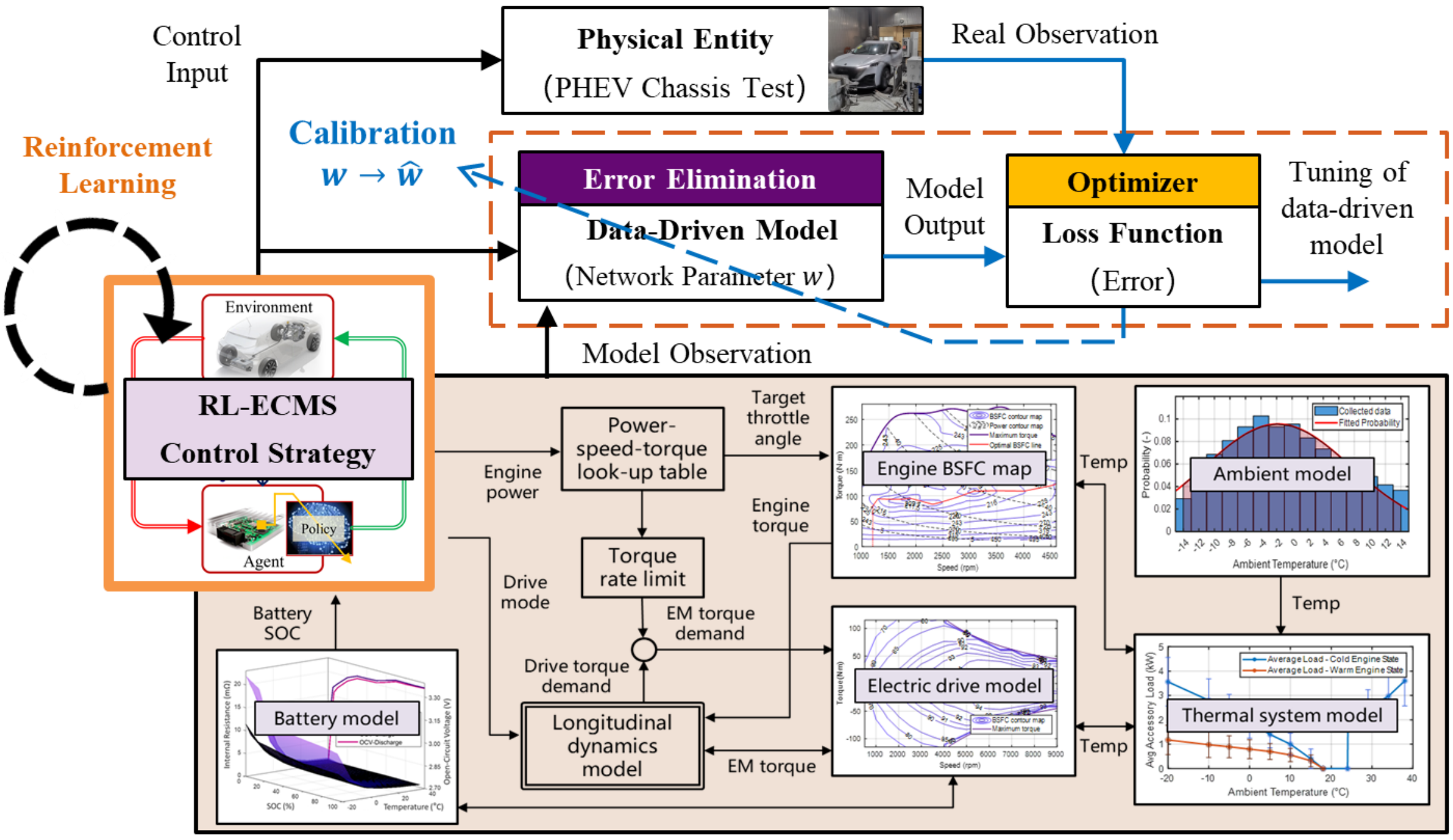

Fig. 7. Data-driven model error calibration approach for modeling fidelity enhancing and RL agent training.

This work adopts the long short-term memory (LSTM) neural network to establish the data-driven error correction model. During training, the error between $\dot{m}_{\mathrm{fuel}}^{\mathrm{a}}$ and $\dot{m}_{\mathrm{fuel}}^{\mathrm{r}}$ is defined as the training loss, with LSTM output denoted as $\Delta\dot{m}_{\mathrm{fuel}}$. To analyze the advantages of LSTM, this study also compares it with other data-driven methods, including support vector machine (SVM), multi-layer perceptron (MLP) model and recurrent neural network (RNN) model, using the same inputs and outputs as the LSTM model. The principle of LSTM cells and gating mechanisms are summarized in Eq. (12), and the detailed principle can be referred to the authors' previous papers [X,X].

$$\begin{cases} i(t) = \sigma\left(W_{x,i}x(t) + W_{h,i}h(t-1) + W_{c,i}c(t-1) + b_i\right) \\ f(t) = \sigma\left(W_{x,f}x(t) + W_{h,f}h(t-1) + W_{c,f}c(t-1) + b_f\right) \\ c(t) = f(t) \cdot c(t-1) + i(t) \cdot tanh\left(W_{x,c}x(t) + W_{h,c}h(t-1) + b_c\right) \\ o(t) = \sigma\left(W_{x,o}x(t) + W_{h,o}h(t-1) + W_{c,o}c(t) + b_o\right) \\ h(t) = o(t) \cdot tanh\left(c(t)\right) \end{cases} \tag{12}$$

where $i$, $f$, and $o$ represent the states of the input gate, forget gate, and output gate, respectively. Additionally, $c$ and $h$ represent the cell memory state and output vector. $W_{x,i}$, $W_{h,i}$, $W_{c,i}$, $W_{x,f}$, $W_{h,f}$, $W_{c,f}$, $W_{x,c}$, $W_{h,c}$, $W_{x,o}$, $W_{h,o}$, and $W_{c,o}$ are linear transformation matrices, and $b_i$, $b_f$, $b_c$, and $b_o$ are bias vectors. $\sigma$ and tanh respectively represent the sigmoid function and the hyperbolic tangent function.

In the design and training process of the data-driven model, feature extraction is necessary, which eliminates irrelevant variables. This study uses a feature selection method combining eXtreme Gradient Boosting (XGBoost) with Pearson correlation analysis. XGBoost is a machine learning algorithm based on gradient-boosted decision trees, classifying and regressing tasks by integrating multiple decision tree models. It iteratively trains decision tree models using a gradient boosting algorithm, thereby gradually improving model performance. The algorithm's objective function is shown in Eq. (13).

$$Obj_{\mathrm{X}}(t) = \sum_{i=1}^{n} l\left(y_i, \hat{y}_i(t-1) + f_t(x_i)\right) + \Omega(f_t) + \mathrm{const} \tag{13}$$

where $l$ is the loss function, $\Omega$ represents the regularization term, and $const$ indicates a constant term. After XGBoost feature selection, the importance weights of the variables are shown in Fig. 8 (a). Results indicate that $\dot{m}_{\mathrm{ac}}$ and $p_{\mathrm{exh}}$ appear more often in decision tree splitting with the XGBoost algorithm, highlighting their significance for this model. Other relevant variables include $\theta$, $T_{\mathrm{ICE}}^{\mathrm{cmd}}$, $T_{\mathrm{ICE}}^{\mathrm{c}}$, $\dot{m}_{\mathrm{EGR}}$, $EGR_{\mathrm{vp}}$, $\omega_{\mathrm{ICE}}$, $\dot{m}_{\mathrm{at}}$, and $VT$. The five least important variables are $\eta_{\mathrm{vol}}$, $p_{\mathrm{int}}$, $\lambda$, $p_{\mathrm{oil}}$, and $T_{\mathrm{int}}$, whose effects are ignored in subsequent modeling.

The Pearson correlation coefficient is used to measure the degree of correlation among different variables. Its value ranges from -1 to 1, with a larger absolute value indicating a stronger correlation. Eq. (14) shows the expression for the Pearson correlation coefficient $\rho$, where $x_i$ and $y_i$ respectively represent the independent variable and the dependent variable. The correlation matrices after XGBoost feature selection are shown in Fig. 8 (b) and Fig. 8 (c). Here, $p_{\mathrm{exh}}$ and $T_{\mathrm{ICE}}^{\mathrm{cmd}}$ scored high on XGBoost importance but had high correlation coefficients with other variables. After XGBoost-Pearson feature selection, the input variables for the data-driven model are determined to be $\dot{m}_{\mathrm{ac}}$, $\theta$, $T_{\mathrm{ICE}}^{\mathrm{c}}$, $\dot{m}_{\mathrm{EGR}}$, $EGR_{\mathrm{vp}}$, and $\omega_{\mathrm{ICE}}$, as shown in Fig. 8 (d).

$$\rho = \frac{\sum_{i=1}^{n} x_i y_i - \frac{\sum_{i=1}^{n} x_i \sum_{i=1}^{n} y_i}{n}}{\sqrt{\sum_{i=1}^{n} x_i^2 - \frac{(\sum_{i=1}^{n} x_i)^2}{n}} \cdot \sqrt{\sum_{i=1}^{n} y_i^2 - \frac{(\sum_{i=1}^{n} y_i)^2}{n}}} \tag{14}$$

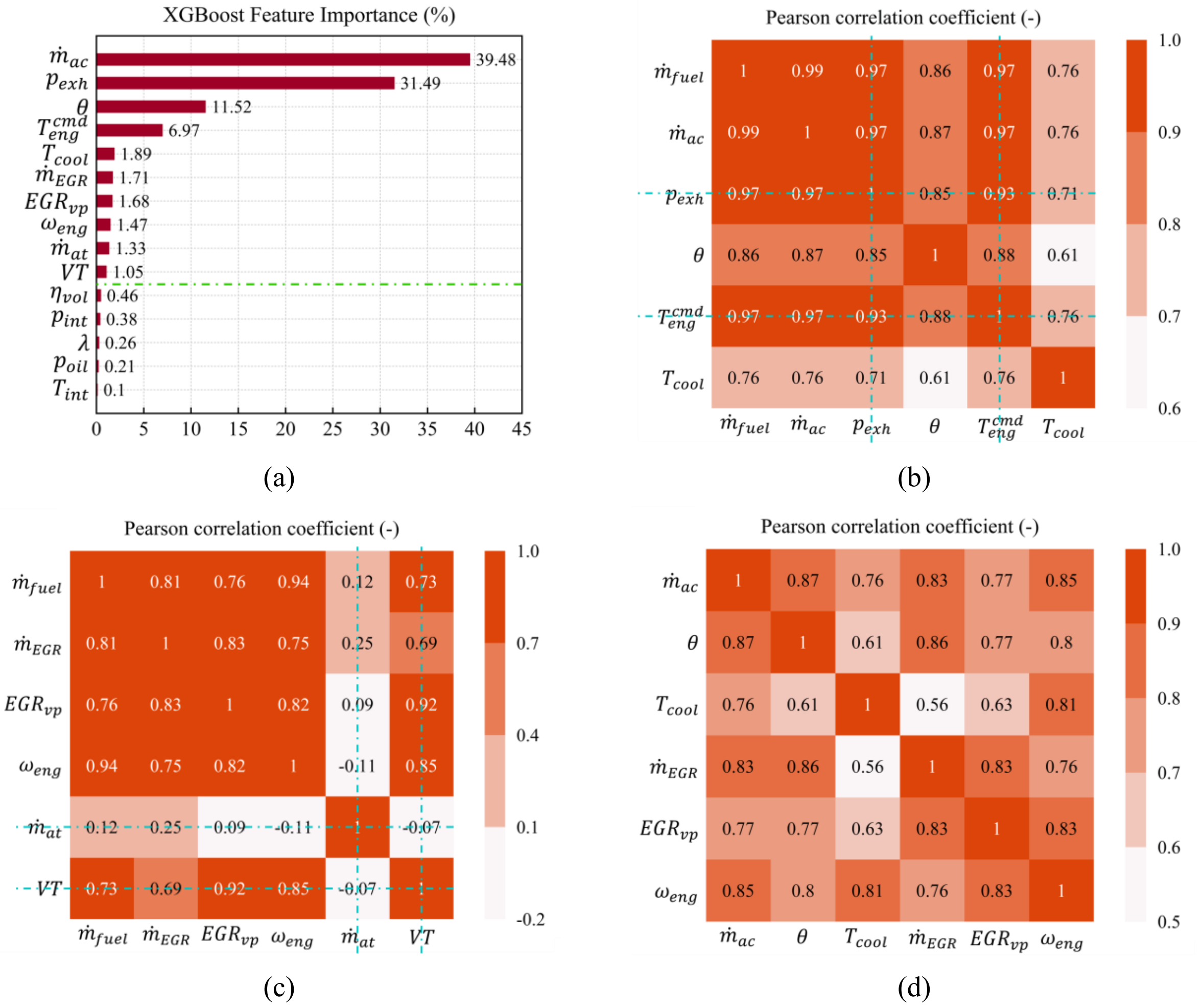

Fig. 8. Results of XGBoost feature importance (a), feature selection using Pearson correlation analysis (b) (c) and the Pearson correlation profile of the selected features (d).

From the feature extraction, it was found that $T_{\text{ICE}}^{\text{c}}$ has a significant impact on fuel consumption. Thus, a high-precision model for $T_{\text{ICE}}^{\text{c}}$ can be established based on data collected from hub tests to improve the accuracy of the fuel consumption correction model. After the same feature selection process, this model uses $\omega_{\text{ICE}}$, $T_{\text{int}}$, $T_{\text{EGR}}$, $T_{\text{oil}}$, and $T_{\text{ICE}}^{\text{cd}}$ as inputs to fit the error $\Delta T_{\text{ICE}}^{\text{c}}$ between experiment and simulation coolant temperatures, as shown in Eq. (15) and Eq. (16).

$$\begin{cases} X_1 = [\omega_{\text{ICE}}, T_{\text{int}}, T_{\text{EGR}}, T_{\text{oil}}, T_{\text{ICE}}^{\text{cd}}] \\ Y_1 = \Delta T_{\text{ICE}}^{\text{c}} = T_{\text{ICE}}^{\text{cr}} - T_{\text{ICE}}^{\text{cd}} \\ \hat{Y}_1(n) = \Delta \hat{T}_{\text{ICE}}^{\text{c}}(n) = LSTM\{[X_1, Y_1]^T(n-M), \dots, X_1^T(n)\} \end{cases} \tag{15}$$

$$T_{\text{ICE}}^{\text{c}} = T_{\text{ICE}}^{\text{cd}} + \Delta \hat{T}_{\text{ICE}}^{\text{c}} \tag{16}$$

where $M$ represents the temporal length of historical data in the LSTM model. The fuel consumption $\dot{m}_{\text{fuel}}$ output by the mechanism-data hybrid-driven model is obtained according to Eq. (17) and Eq. (18). To meet practical requirements, the LSTM network established in this paper has a single input layer and LSTM layer, with the engine water temperature model having an input dimension of 5 and an output dimension of 1, and the engine fuel consumption model having an input dimension of 6 and an output dimension of 1.

$$\begin{cases} X_2 = \left[\dot{m}_{\mathrm{ac}}, \theta, T_{\mathrm{ICE}}^{\mathrm{c}}, \dot{m}_{\mathrm{EGR}}, EGR_{\mathrm{vp}}, \omega_{\mathrm{ICE}}\right] \\ Y_2 = \Delta\dot{m}_{\mathrm{fuel}} = \dot{m}_{\mathrm{fuel}}^{\mathrm{r}} - \dot{m}_{\mathrm{fuel}}^{\mathrm{a}} \\ \hat{Y}_2(n) = \Delta\hat{\dot{m}}_{\mathrm{fuel}}(n) = LSTM\{[X_2, Y_2]^T(n-M), \dots, X_2^T(n)\} \end{cases} \tag{17}$$

$$\dot{m}_{\mathrm{fuel}} = \dot{m}_{\mathrm{fuel}}^{\mathrm{a}} + \Delta\hat{\dot{m}}_{\mathrm{fuel}} \tag{18}$$

*3.2. Integrating horizon-extended learning with real-time optimization framework*

Before integrating RL with the real-time control framework, the ECMS should be briefly introduced. The main idea of ECMS is to use the EF to convert the electrical energy consumption into fuel consumption. Eq. (19) describes the objective function, where the power demand at each time step is defined as $P_{\mathrm{dem}}$, the initial SOC is $SOC_0$, and the optimization problem is subject not only to the physical constraints of component speed, torque, and power but also to the allowable SOC tracking constraint $\varepsilon$. This paper chooses battery output power $P_{\mathrm{bat}}$ as the control variable, and the engine's output power $P_{\mathrm{ICE}}$ can be uniquely determined by $P_{\mathrm{dem}}$ and $P_{\mathrm{bat}}$, thereby limiting the number of control variables to one to reduce the computational burden. Therefore, the engine's fuel consumption rate $\dot{m}_{\mathrm{fuel}}$ at each moment is a function of $P_{\mathrm{bat}}$, where the engine's torque and speed can be uniquely determined by the optimal operating line (OOL).

$$J = \int_{t_0}^{t_{\mathrm{f}}} \dot{m}_{\mathrm{fuel}}\big(P_{\mathrm{bat}}(t)\big)dt$$
$$\mathrm{s.t.} \begin{cases} SOC(0) = SOC_0, SOC(t_{\mathrm{f}}) - SOC_{\mathrm{ref}}(t_{\mathrm{f}}) < \varepsilon \\ P_{\mathrm{dem}}(t) = P_{\mathrm{bat}}(t) + P_{\mathrm{ICE}}(t) \\ P_{\min}^{\mathrm{bat}} \le P_{\mathrm{bat}}(t) \le P_{\max}^{\mathrm{bat}} \\ SOC_{\min} \le SOC(t) \le SOC_{\max} \end{cases} \tag{19}$$

Further, to ensure the battery SOC always remains within a reasonable range, the ECMS problem's Hamiltonian function $H$ is defined as in Eq. (20):

$$\begin{aligned} P_{\mathrm{bat}}^*(t) &= \underset{P_{\mathrm{bat}}}{\operatorname{argmin}}\, H\big(P_{\mathrm{bat}}(t), p(t)\big) \\ &= \underset{P_{\mathrm{bat}}}{\operatorname{argmin}} \Big(\dot{m}_{\mathrm{f}}\big(P_{\mathrm{bat}}(t)\big) - p(t)\dot{SOC}\big(P_{\mathrm{bat}}(t)\big)\Big) \end{aligned} \tag{20}$$

where $p$ is the co-state variable in the PMP problem, $\dot{SOC}$ represents the system's state equation. Furthermore, $\dot{SOC}$ can be rewritten as shown in Eq. (21):

$$\dot{SOC}\big(P_{\mathrm{bat}}(t)\big) = -\frac{100\%}{3600C}\frac{U_{\mathrm{OC}} - \sqrt{{U_{\mathrm{OC}}}^2 - 4R_{\mathrm{bat}}P_{\mathrm{bat}}(t)}}{2R_{\mathrm{bat}}} \tag{21}$$

The ECMS equivalence factor $\lambda$ can be defined as in Eq. (22), where $\tilde{\eta}_{\mathrm{ICE}}$ is defined as an estimate of the engine's comprehensive efficiency under actual driving conditions, a constant typically set to 50% of the engine's peak thermal efficiency. The Hamiltonian function $H$ represented by the equivalence factor $\lambda$ is as shown in Eq. (23).

$$\lambda(t) = -\frac{100\%\mathrm{LHV}\tilde{\eta}_{\mathrm{ICE}}}{3600CU_{\mathrm{OC}}}p(t) \tag{22}$$

$$\begin{aligned} H\big(P_{\mathrm{bat}}(t),\lambda(t)\big) &= \dot{m}_{\mathrm{f}}\big(P_{\mathrm{bat}}(t)\big) - p(t)S\dot{O}C\big(P_{\mathrm{bat}}(t)\big) \\ &= \dot{m}_{\mathrm{f}}\big(P_{\mathrm{bat}}(t)\big) + \lambda(t)\frac{P_{\mathrm{bat}}(t)}{\mathrm{LHV}} \end{aligned} \tag{23}$$

By regarding the hybrid vehicle with ECMS as environment $E$, and the equivalence factor controller as the RL agent, thus modeling the energy management problem as an MDP and proposing the RL-based control architecture, namely RL-ECMS, as shown in Fig. 9. This problem includes several core elements: states $S$, actions $A$, state transition probability $P$, and reward function $R$, forming the tuple $\{S, A, P, R\}$. Through interaction with the environment, the RL agent learns how to formulate strategies aimed at maximizing its long-term cumulative reward for energy management. When the agent takes an action $a \in A$, i.e., outputs the evaluation value of the equivalence factor, and acts on the current state $s$, the underlying ECMS controller will determine a unique power distribution relationship. Then, the powertrain will switch from state $s$ to another state $s'$ according to the state transition probability $P$, while the reward function $R$ provides a reward feedback to the agent.

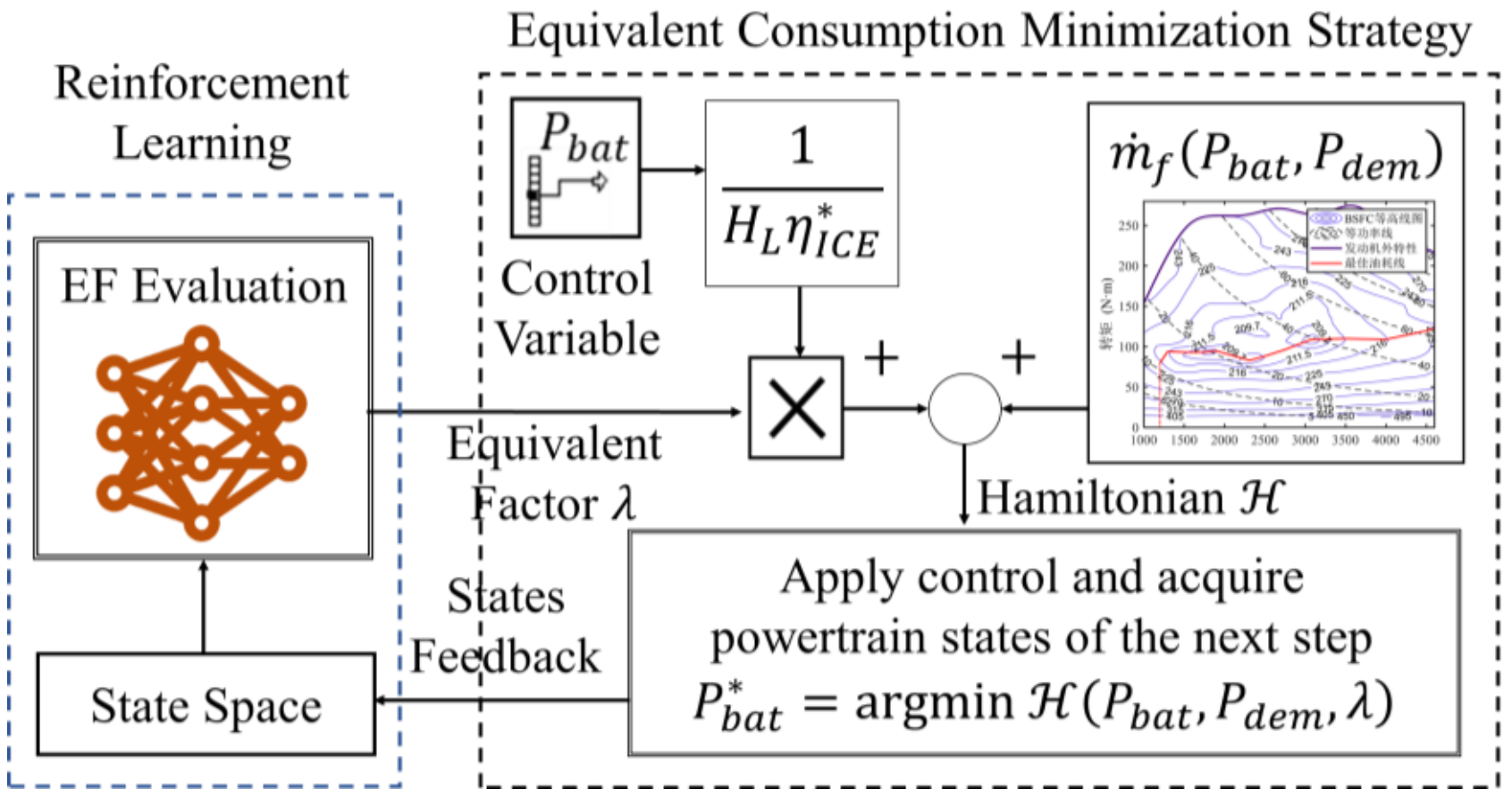

Fig. 9. The RL-ECMS coupling framework based on learning-based equivalent factor evaluation.

The horizon-extended reinforcement learning is adopted for equivalent factor evaluation, where the state space includes the vehicle's average demand power $P_{\mathrm{avg}}$ over the past horizon $T_{\mathrm{avg}}$, the estimated vehicle speed $V_{\mathrm{FX}}$ within an extended future horizon $T_{\mathrm{FX}}$, which can be obtained from digital maps in real application. Moreover, the states also include the current moment engine output power $P_{\mathrm{ICE}}$, current vehicle speed $v$, distance to the destination $d_{\mathrm{rem}}$, and the difference between actual SOC and target SOC $\Delta SOC$, as shown in Eq. (24):

$$s = \left[P_{\text{avg}}, V_{\text{FX}}(t\colon t + T_{\text{FX}}), P_{\text{ICE}}, v, d_{\text{rem}}, \Delta SOC\right] \quad (24)$$

The agent evaluates the optimal equivalence factor $\lambda$, as defined in Eq. (25), according to the policy $\pi_\theta(s)$. The range of action variable $a$ is determined based on simulation results. Extensive global optimization of the equivalence factor under various standard cycle conditions has revealed that the range of the equivalence factor is consistently between 1 and 1.5. This implies that the overall fuel consumption is 1 to 1.5 times that of the engine peak thermal efficiency due to the inevitable efficiency losses led by the powertrain and complex driving cycles. Hence, the output range of the equivalence factor is chosen to be 0.5 to 2, encompassing the rational operating range of the equivalence factor with sufficient margin:

$$a = \{\lambda \mid 0.5 \le \lambda \le 2\} \quad (25)$$

The reward function $r$ includes the engine fuel consumption rate $\dot{m}_{\text{fuel}}$, in L/s, and the battery SOC penalty $\varsigma$, both set as negative values, as shown in Eq. (25). The adjustment coefficient $\tau$ balances engine fuel consumption and battery energy consumption. When $\Delta SOC$ is greater than zero, i.e., the actual vehicle SOC is greater than the target SOC value, no penalty for SOC is needed because a higher SOC already corresponds to more fuel consumption. Conversely, when $\Delta SOC$ is less than zero, i.e., the actual vehicle SOC is below the target SOC value, a penalty for the squared deviation value $\Delta SOC$ is activated to prevent the agent from trading excessive discharge for lower fuel consumption, thus ensuring the battery SOC's operating range is safe and controllable:

$$\begin{gathered} r(s,a) = -[\dot{m}_{\text{fuel}}(a) + \tau\varsigma(s)] \\ \varsigma(s) = \begin{cases} 0, & \Delta SOC \ge 0 \\ \Delta SOC^2, & \Delta SOC < 0 \end{cases} \end{gathered} \quad (25)$$

When the agent takes action $a$ based on state $s$ at time $t$, it receives a reward signal $r(s,a)$ as shown in Fig. 10. The optimal equivalence factor evaluation policy $\pi^*$ that maximizes cumulative returns is defined as shown in Eq. (26) to improve the long-term overall fuel economy of the powertrain and maintain the battery SOC at a certain level.

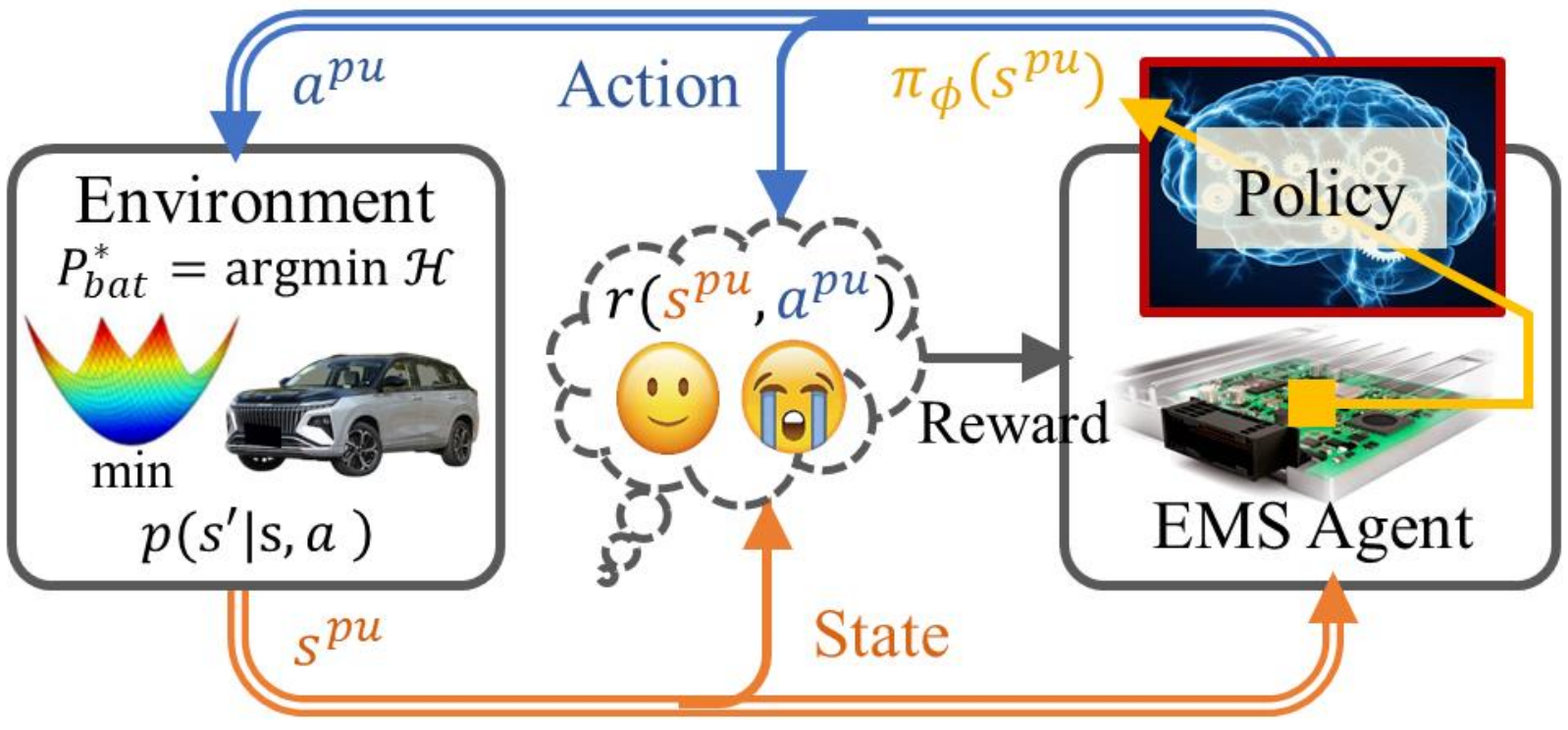

Fig. 10. Interaction between RL agent and the hybrid powertrain based on the ECMS control framework.

$$\pi^* = \underset{\pi}{\operatorname{argmax}} V^\pi(s) = \underset{\pi}{\operatorname{argmax}} \mathbb{E}_\pi \left[ \sum_{t=1}^{T} \gamma^{t-1} r(s,a) \right] \tag{26}$$

As shown in Fig. 11, a randomly initialized Actor network with exploratory noise is used to output continuous equivalence factor, where $\theta$ represents the network parameters, while Gaussian noise $\varepsilon$, following a $N(0,\delta)$ distribution, is added to the deterministic policy to balance exploration and exploitation. This iteration is repeatedly performed to collect experience tuples $e_t = (s_t, a_t, r_t, s_{t+1})$. The paper employs two evaluate Critic networks to compute Q-values based on state and action inputs in parallel to reduce the overestimation issue, denoted as $Q_{w_1}(s_t,a_t)$ and $Q_{w_2}(s_t,a_t)$. Additionally, $Q_{w_1'}(s_{t+1},\tilde{a}_{t+1})$ and $Q_{w_2'}(s_{t+1},\tilde{a}_{t+1})$ are defined as the Q-values computed by the two target Critic networks. Where $\tilde{a}_t$ represents the action generated by the target Actor network, which, after adding truncated random noise, is used for smoothing the state-action value estimation. The smaller of the two Q-values output by the target Critic networks is used to minimize the loss function based on temporal-difference (TD) error, thus updating the Critic network's parameters, as shown in Eq. (27) to Eq. (29).

$$J(w_i) = \frac{1}{M} \sum_{t=1}^{M} \left[ y_t - Q_{w_i}(s_t,a_t) \right]^2, \qquad i = 1,2 \tag{27}$$

$$y_t = r_t + \gamma \min\left[ Q_{w_1'}(s_{t+1},\tilde{a}_{t+1}), Q_{w_2'}(s_{t+1},\tilde{a}_{t+1}) \right] \tag{28}$$

$$\tilde{a}_{t+1} = \pi_{\theta'}(s_{t+1}) + clip[N(0,\delta'),-c,c] \tag{29}$$

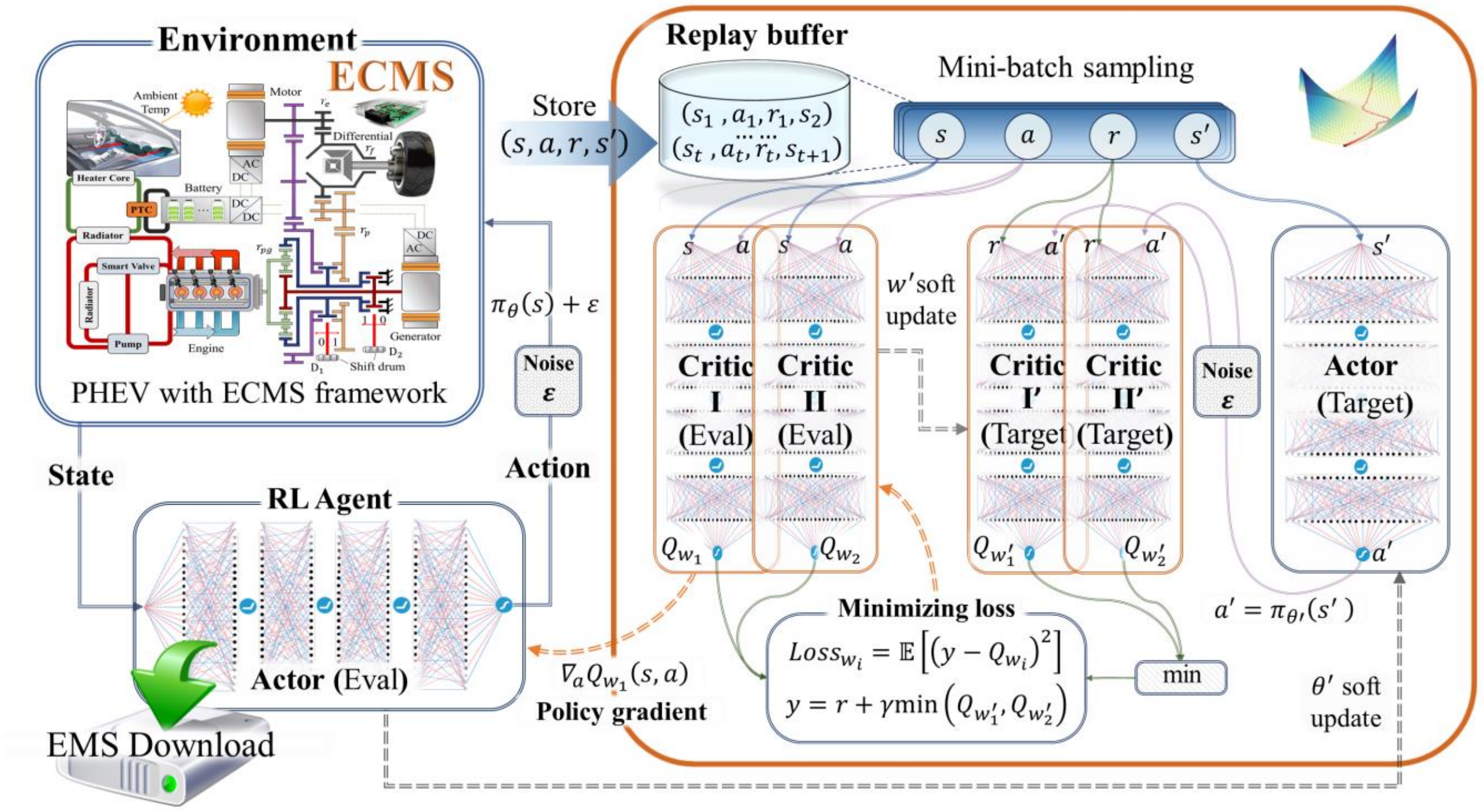

Fig. 11. Training framework for RL-ECMS energy management based on TD3 algorithm.

For each experience tuple $e_t = (s_t, a_t, r_t, s_{t+1})$, $s_t$ is input into the evaluate Actor network to generate action. The obtained action is then input into two evaluate Critic networks for Q-value calculation, denoted as $Q_{w_1}(s_t, \pi_\theta(s_t))$ and $Q_{w_2}(s_t, \pi_\theta(s_t))$, where $w_1$ and $w_2$ are the parameters of the evaluate Critic network I and II, respectively. Moreover, the Q-value $Q_{w_1}(s_t, \pi_\theta(s_t))$ generated by evaluate Critic network I is used to update the parameters of the Actor network using the deterministic policy gradient according to Eq. (30). It's noteworthy that the updates of the Critic and Actor networks are performed alternately, and the target value networks are updated after every $d$ steps, meaning that the target value networks' weights are updated slower than the current value networks.

$$\begin{aligned} \nabla_\theta J(\theta) &= \mathbb{E}\left[\nabla_a Q_{w_1}(s_t, a)|_{a=\pi_\theta(s_t)} \nabla_\theta \pi_\theta(s_t)\right] \\ &\approx \frac{1}{M}\sum_{m=1}^{M} \nabla_a Q_{w_1}(s_j, a)|_{a=\pi_\theta(s_j)} \nabla_\theta \pi_\theta(s_j) \end{aligned} \tag{30}$$

Before the training of the energy management strategy based on TD3, the capacity of the experience replay buffer $D$ is initialized to $M$. During the interaction process, the agent will iterate repeatedly and place new experience tuples into the replay buffer $D$, and when the buffer capacity is full, the earliest experience tuples will be replaced by new ones. During the training process, a batch of $N_s$ experience tuples is randomly sampled from the replay buffer for network updates. Furthermore, the parameters of the target networks inherit from the corresponding evaluate networks through soft updates, where the soft update ratio can be customized.

## 4. Testing and validation setup

### *4.1. Validation setup*

To validate the real-time control performance of the proposed RL-ECMS strategy, this work conducts both simulation and hardware-in-the-loop (HIL) tests. In the HIL tests, this work sets the initial coolant temperature to be 90 °C, ignoring the cold start impacts. The initial SOC and target SOC are both set as 34%, consistent with the OEM's benchmark strategy. Fig. 12 shows the configuration of the HIL testing platform, including a real-time personal computer (RTPC) of the National Instruments (NI) PXIe-1078 series, used for real-time simulation of vehicle dynamics, equipped with the PXI-6221 multifunction analog acquisition card. The rapid control prototype (RCP) acts as the energy management controller, allowing for real-time testing of control strategies. Additionally, the HIL upper computer is used to download HIL vehicle models, coordinate the simulation environment, and execute test scripts. The RCP master computer downloads the energy management strategy to the RCP controller via controller area network (CAN) communication. The Input/Output (I/O) signals of the RCP controller are connected to the RTPC via CAN communication, consistent with the real vehicle application.

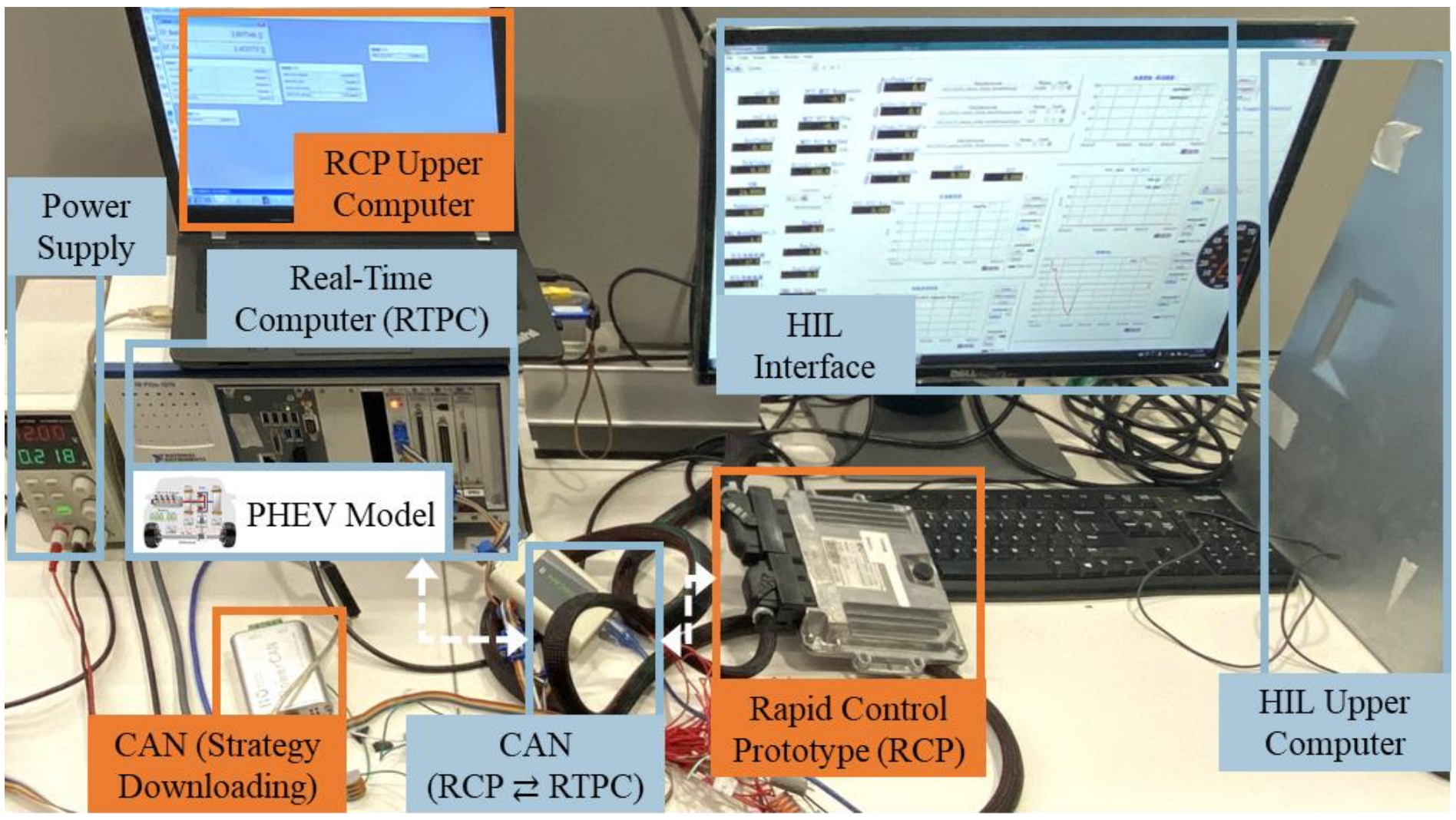

Fig. 12. Hardware-in-the-loop test setup for real-time applicability and optimality validation of RL-ECMS.

### *4.2. Baseline Strategies*

This paper compares the RL-ECMS energy management strategy trained under the NEDC standard cycle with the following four strategies: an offline optimal control strategy based on dynamic programming, a traditional RL energy management strategy with engine power output as the action, an A-ECMS energy management system, and a rule-based benchmark energy management strategy provided by OEM. Moreover, the traditional reinforcement learning energy management strategy is denoted as the RL strategy in the following comparisons, with its state space as $[v, a, \Delta SOC]$, corresponding to vehicle speed, acceleration, and battery SOC difference, respectively, and the action directly set as engine power $P_{\mathrm{ICE}}$, with the training method also using the TD3 algorithm to ensure fairness of comparison. Additionally, for A-ECMS, its equivalence factor initial value and PI controller parameters are optimized to ensure fairness of comparison, with initial values chosen in the range of 0.5~2 to remain consistent with RL-ECMS, and the PI controller's proportional and integral coefficients optimized within ranges of 2~15 and 0.05~0.5, respectively. For fuel economy comparison, this paper, according to SAE J1711 standard, corrects the final SOC under different control strategies to the same value to evaluate the converted fuel consumption, ensuring fairness of fuel economy comparison.

## 5. Results and discussions

Fig. 13 compares the effectiveness of different engine coolant temperature models in the training and validation sets, using the mean absolute error (MAE) metric. The scatter plot in Fig. 13 (a) represents the data distribution of various models during the training process. The LSTM model performed best in fitting the experimental and simulation data errors. Fig. 13 (b) describes the error distribution between test values and simulation values in the training set in histogram form. For the LSTM-based correction model, its MAE is the smallest among the four methods.

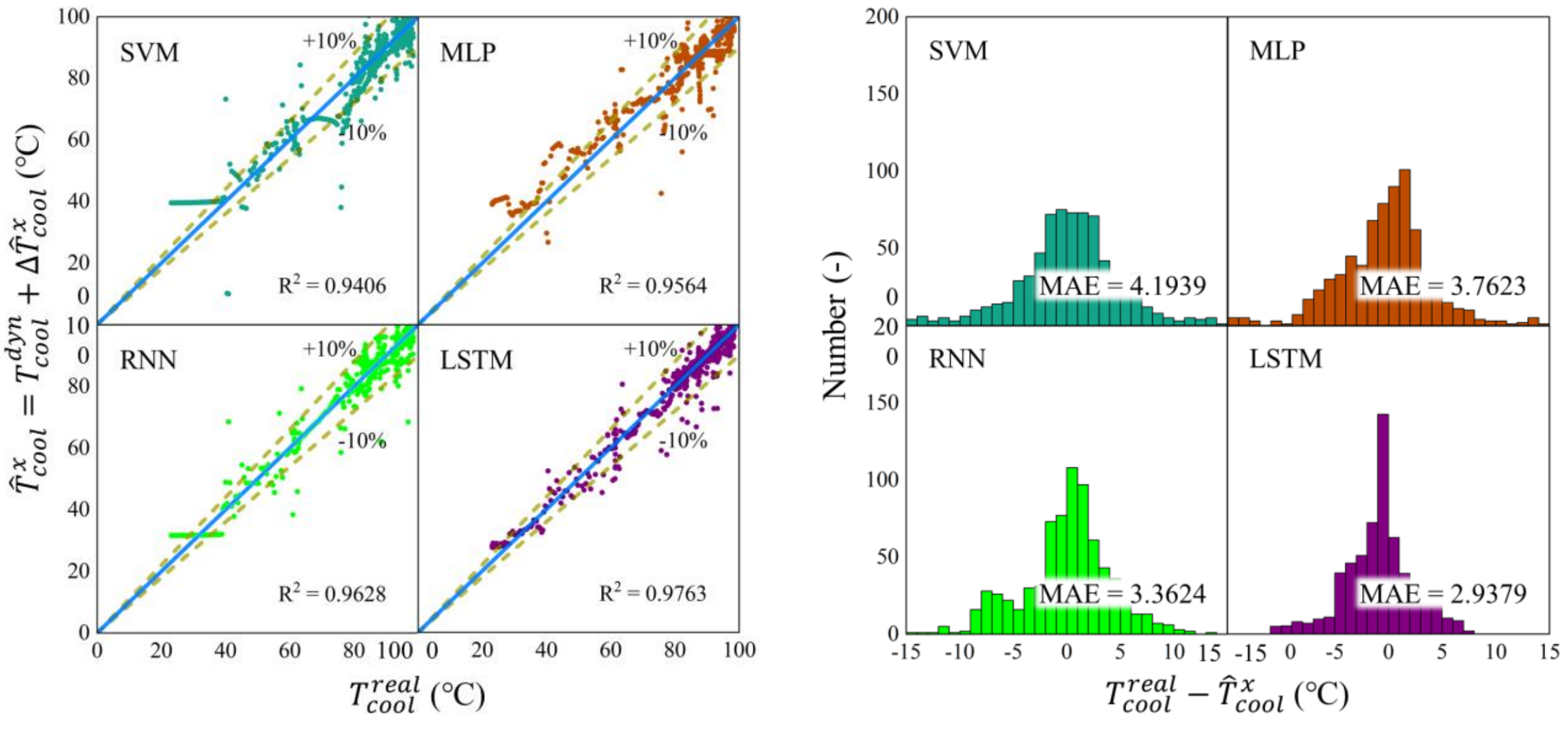

(a) (b)

Fig. 13. Results of the engine coolant temperature modeling in terms of comparison of model output and real data (a) and the model MAE using different machine learning algorithms (b).

The performance of the fuel consumption model in the training and validation sets is analyzed below. The scatter plot in Fig. 14 (a) shows the simulated model fuel consumption values $\dot{m}^{x}_{\text{fuel}}$ and the actual fuel consumption values $\dot{m}^{\text{r}}_{\text{fuel}}$ during the training process. Among the four methods, the LSTM-based model's prediction accuracy was notably higher than the other methods, and Fig. 14 (b) compares the histogram data of error distribution. Furthermore, the four correction model are comparatively analyzed as shown in Table 2.

**Table 2**

Comparison of LSTM-based error correction model against conventional models

| Fuel consumption model | MAE | MSE | MRE |
|---|---|---|---|
| Dynamic model $\dot{m}^{\text{d}}_{\text{fuel}}$ | 0.207 | 0.121 | 0.357 |
| Map-based quasi-static model $\dot{m}^{\text{q}}_{\text{fuel}}$ | 0.180 | 0.073 | 0.238 |
| The average value $\dot{m}^{\text{a}}_{\text{fuel}}$ | 0.179 | 0.071 | 0.275 |
| The data-driven corrected model $\hat{\dot{m}}^{\text{LSTM}}_{\text{fuel}}$ | 0.067 | 0.012 | 0.105 |

To validate the accuracy of the established powertrain model, the simulation results after error correction are compared to the experimental data, as shown in Fig. 15. It can be seen that the simulation results exhibit a high degree of consistency with bench test data. The engine output power and battery's SOC changes nearly coincide in simulation and real vehicle tests. The engine coolant temperature and fuel consumption model closely follows the trend of the corresponding real vehicle data. Furthermore, this work selects the new European driving cycle (NEDC) as the training condition for the strategy and the worldwide light-duty vehicle test cycle (WLTC) cycle as the verification condition. A total of 500 training epochs were set, with the network structure including four fully connected layers containing 200, 150, 100, and 50 neurons, respectively. Moreover, the learning rates for the Actor and Critic networks were set to 0.0001, the discount factor $\gamma$ and exploration probability $\varepsilon$ were set to 0.995 and 0.001, respectively. In addition, the soft update interval $d$ for the target network was set to 4, and the clipping random noise parameter $c$ was set to 0.5. The length of the experience replay buffer was set to 1000000, with a sampling amount for experience replay set to 256, as shown in Table 3. Furthermore, the design of the reward function significantly affects the learning process's convergence ability and performance. The core parameter of the reward function is the coefficient $\tau$, aiming to balance the fuel economy and SOC sustaining.

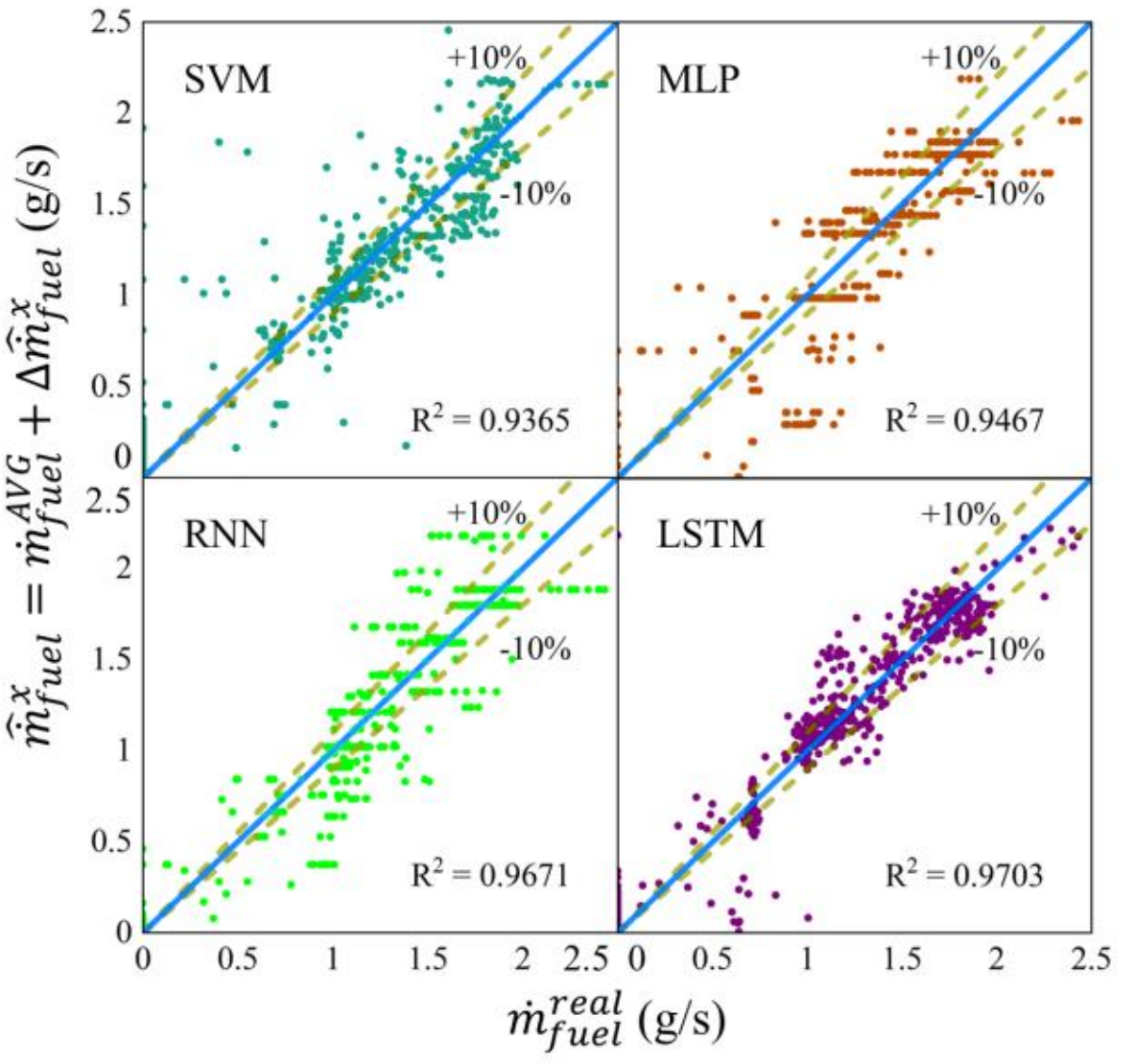

(a)

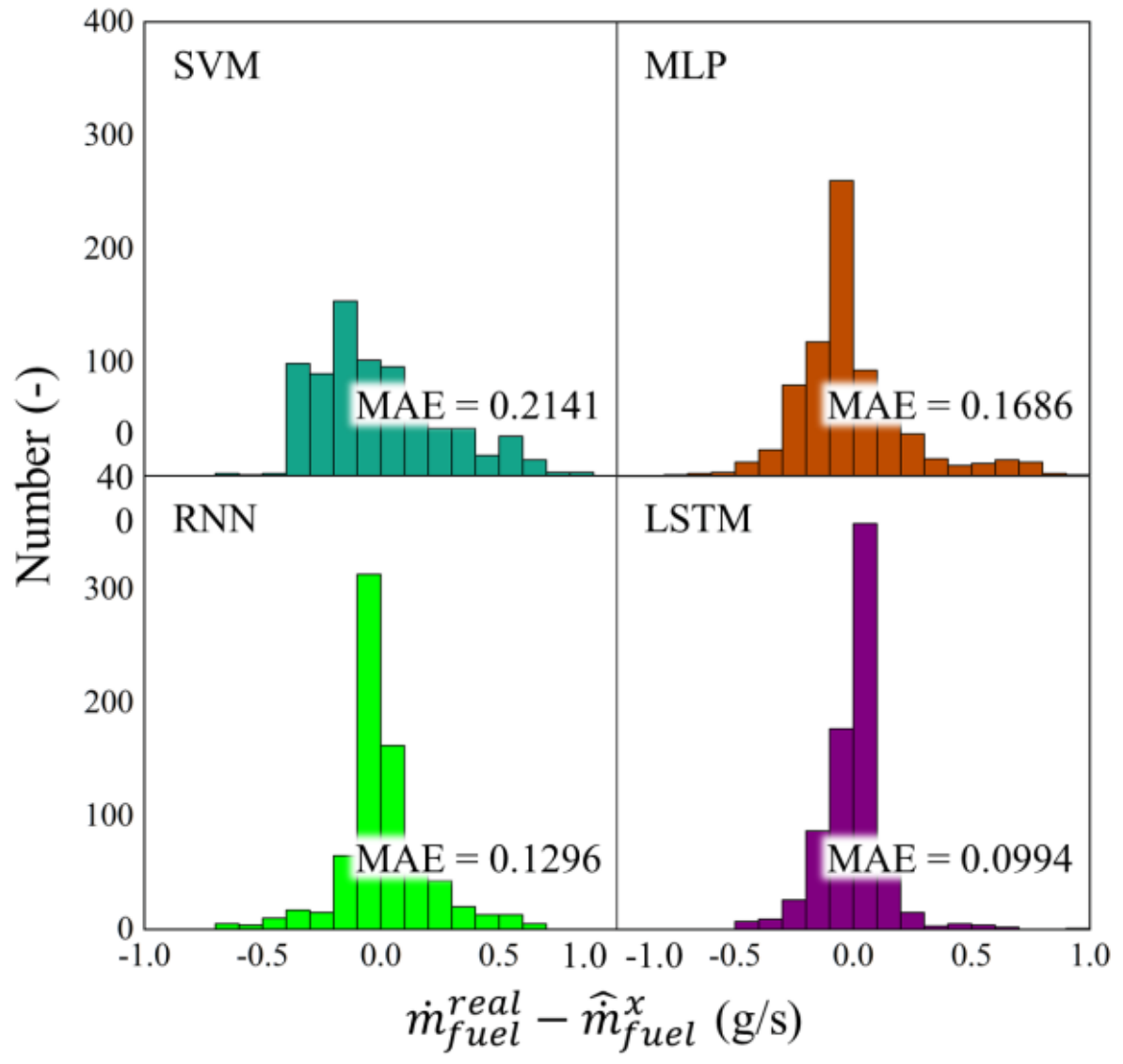

(b)

Fig. 13. Results of the engine fuel consumption modeling in terms of comparison of model output and real data (a) and the model MAE using different machine learning algorithms (b).

The training results of the RL-ECMS with different RL algorithms are shown in Fig. 16, where each episode's return represents the cumulative reward obtained from running a complete driving cycle. The magnitude of the return reflects the overall performance of the energy management agent during the training process. With an increase in the number of training episodes, the cumulative reward first increases and then tends to stabilize. This figure displays the learning process of strategies using different RL algorithms. It is observed that DQN and Double DQN algorithms converge the fastest, completing the training process around 100 episodes. The convergence speeds of PPO and DDPG are slower, completing the training process and converging around the 200th and 250th episodes, respectively. The training speed of the TD3 algorithm falls in between, but its cumulative reward is slightly higher than the other four strategies. The cumulative rewards of Double DQN and DDPG are close but lower than the TD3 algorithm, while DQN and PPO have the lower returns.

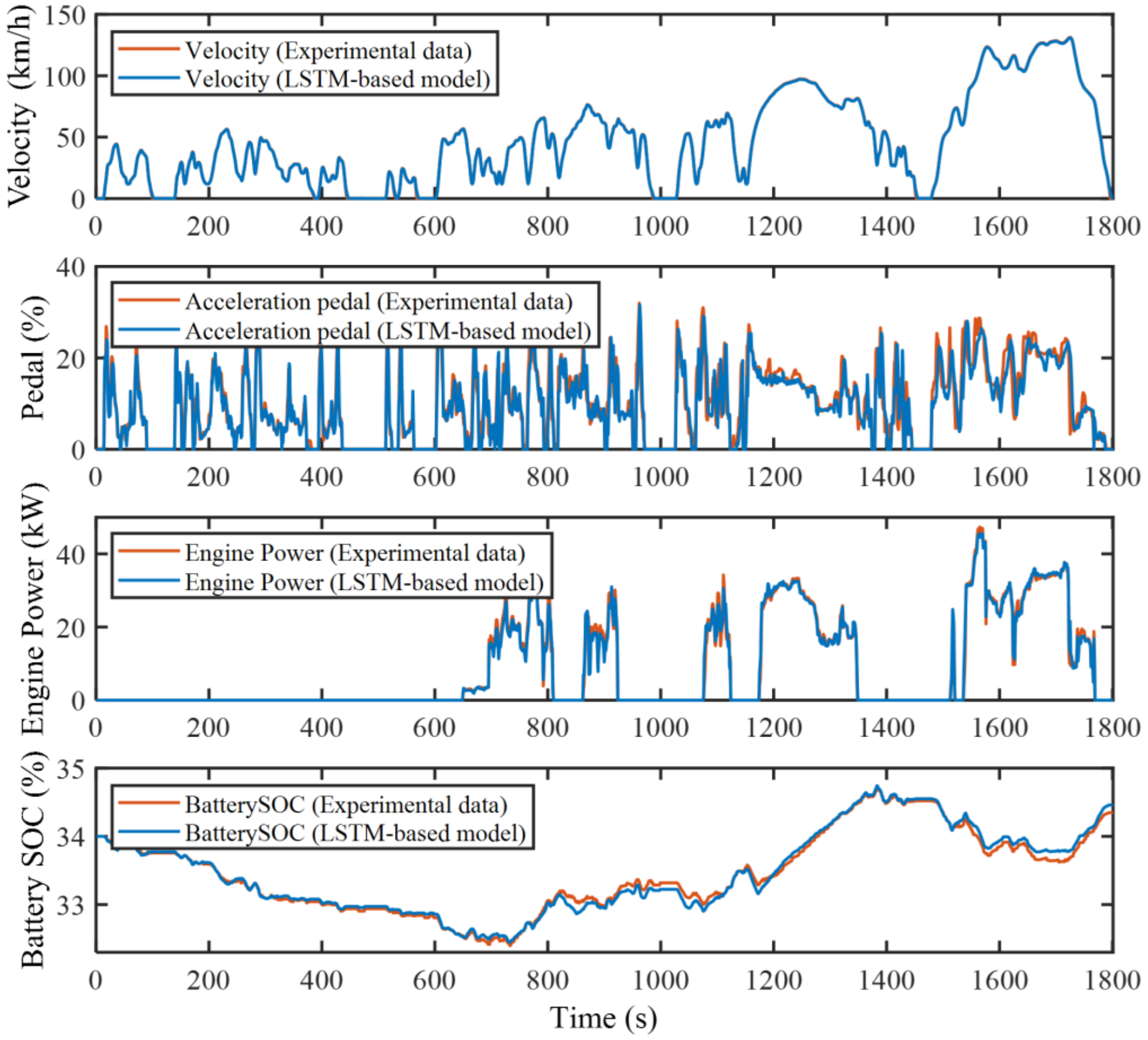

(a)

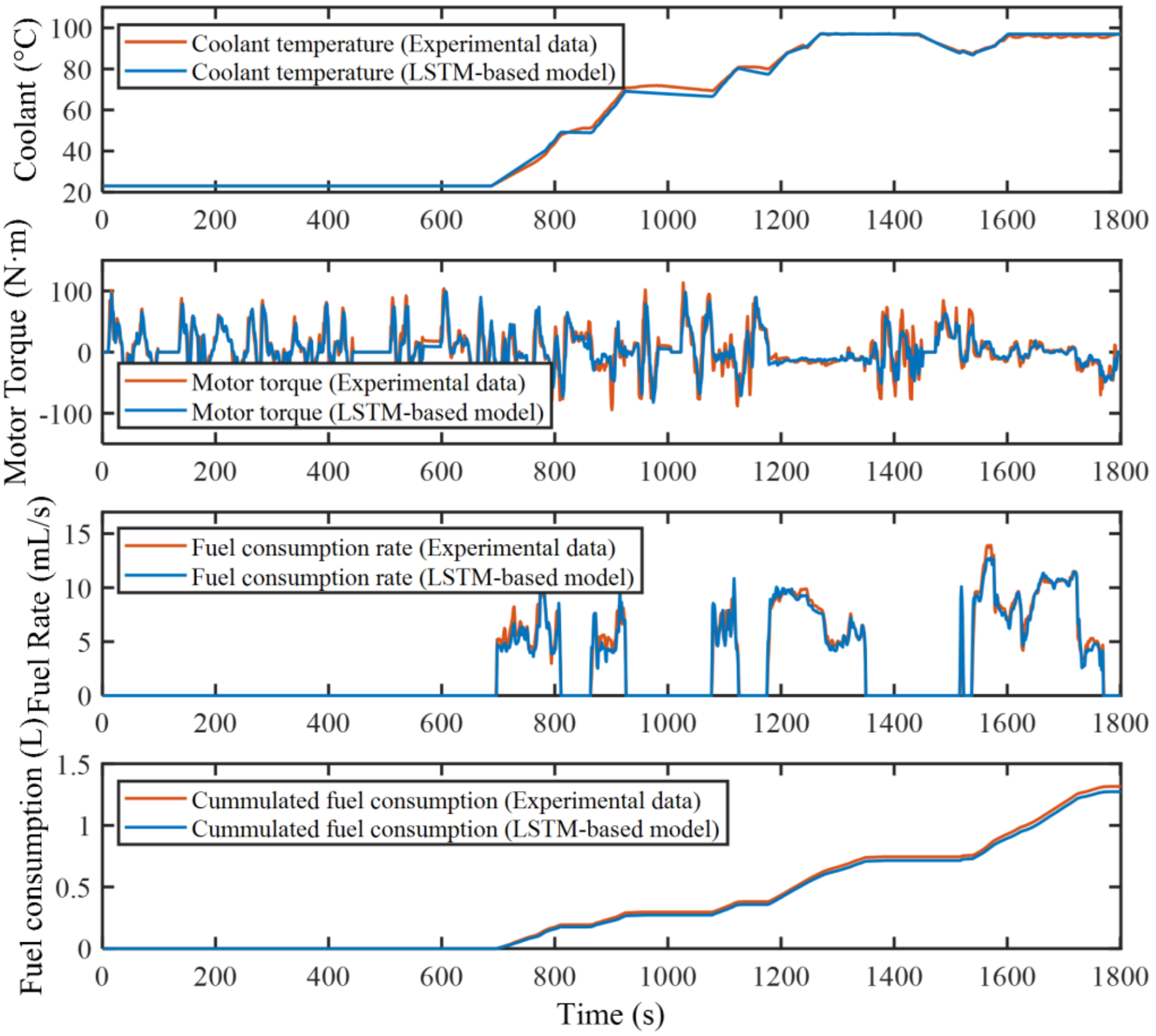

(b)

Fig. 15. Error between the model and experimental results in terms of the driver model, engine power, and battery SOC (a) and coolant temperature, motor torque, fueling rate and cumulative fuel consumption (b).

**Table 3**

Hyperparameters settings for RL-ECMS training

| Hyperparameters | Value | Hyperparameters | Value |
|---|---|---|---|
| Discount factor $\gamma$ | 0.995 | Total episode | 500 |
| Buffer capacity $M$ | 1×106 | Actor learning rate | 1×10-4 |
| Exploration ratio $\varepsilon$ | 0.001 | Critic learning rate | 1×10-4 |
| Batch size $N_s$ | 256 | Soft update rate $\sigma_w$、$\sigma_\theta$ | 0.005 |
| Update delay $d$ | 4 | Noise clip range $c$ | 0.5 |

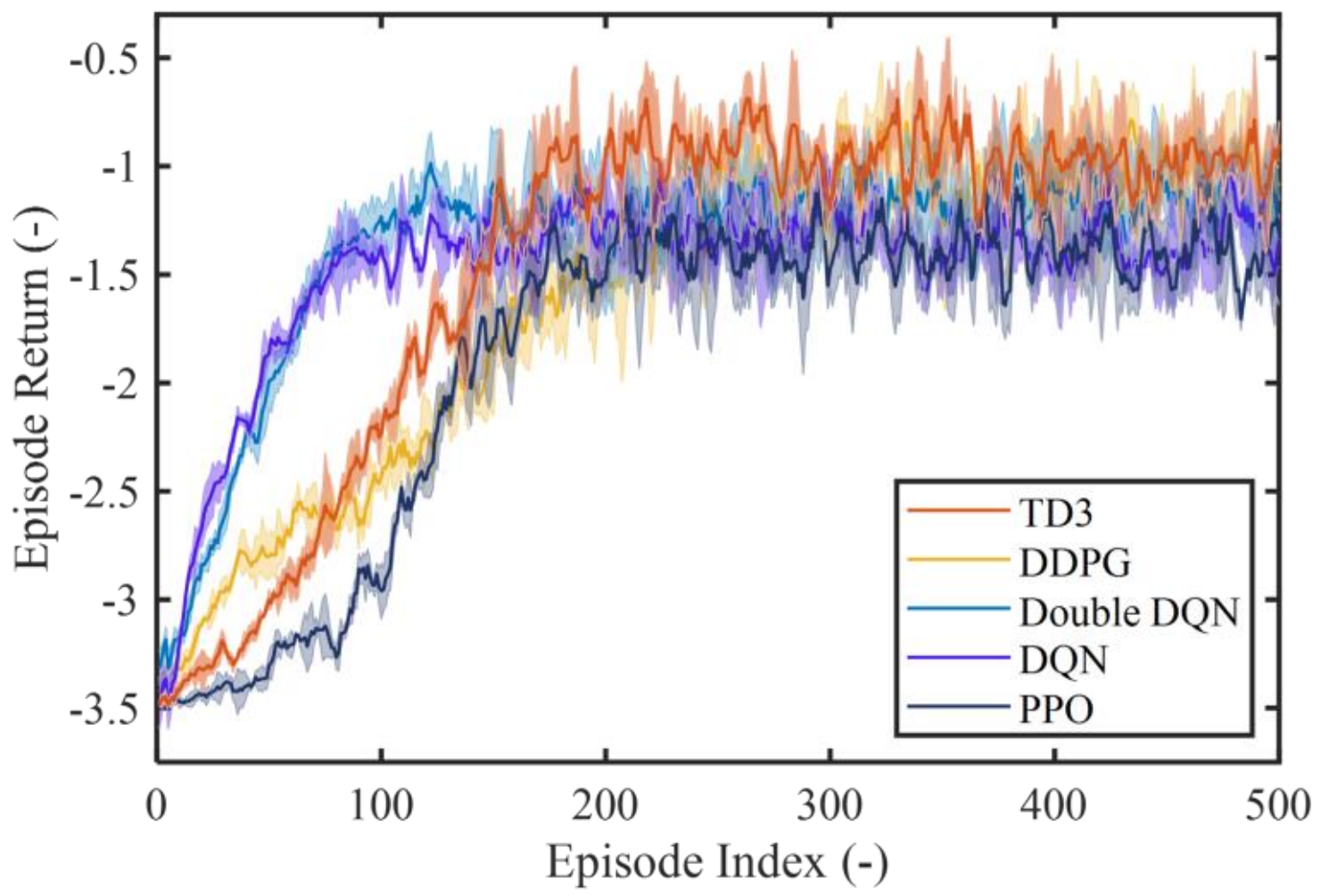

Fig. 16. Comparison of convergence process in different reinforcement learning algorithms.

Moreover, optimization was performed for the parameter $\tau$, and the results are shown in Fig. 17, displaying the control trajectories of the SOC and corresponding fuel consumption under different reward function designs. Table 4 details the control results, it can be seen that with an increase in the value of $\tau$, the final value of battery SOC shows a gradually increasing trend, from 26.86% to 37.88%. This means that as the value of $\tau$ increases, the remaining battery capacity gradually increases, i.e., the energy management strategy tends to preserve electrical energy more. The corrected fuel consumption first slightly decreases from 5.13 L/100 km for $\tau$=1.5 to 4.93 L/100 km for $\tau$=3.0, then gradually increases to 5.15 L/100 km corresponding to $\tau$=4.5. This trend indicates that an initial increase in $\tau$ value enhances the EMS control performance, but as $\tau$ continues to increase, the effect instead decreases.

The following tests are carried out under the WLTC to verify the control performance of the trained RL-ECMS system, and Fig. 18 shows the operation results of different algorithms with statistical results seen in Table 5. It is observable that the engine output power, engine coolant temperature and SOC trajectory of RL-ECMS and RL strategies are similar to the offline optimal solution, consistent with the conclusions under the NEDC. There is a minimal difference in the control results between RL-ECMS and the RL strategy, with corrected fuel consumption at 4.93 L/100 km and 4.96 L/100 km, respectively. Under this condition, the number of engine start-stops for RL-ECMS and the RL strategy significantly increased to 12 and 23 times compared to the NEDC condition, especially with the RL strategy experiencing frequent start-stops under the WLTC condition. In contrast, the A-ECMS strategy and the rule-based benchmark strategy have higher fuel consumption, with corrected fuel consumption at 5.03 L/100 km and 5.17 L/100 km, respectively.

**Table 4**

Comparison of EMS control performance under different reward function design.

| Performance indicator | Parameter design of reward function | | | | | | |
|---|---|---|---|---|---|---|---|
| | $\tau$=1.5 | $\tau$=2.0 | $\tau$=2.5 | $\tau$=3.0 | $\tau$=3.5 | $\tau$=4.0 | $\tau$=4.5 |
| Battery final SOC/ % | 26.86 | 30.42 | 31.88 | 33.90 | 35.84 | 37.16 | 37.88 |
| Fuel consumption/ L | 0.61 | 0.88 | 0.96 | 1.14 | 1.31 | 1.44 | 1.51 |
| Fuel economy/ L/100 km | 2.63 | 3.79 | 4.24 | 4.90 | 5.64 | 6.17 | 6.50 |
| Corrected fuel economy/ L/100 km | 5.13 | 5.04 | 4.98 | 4.93 | 4.99 | 5.07 | 5.15 |

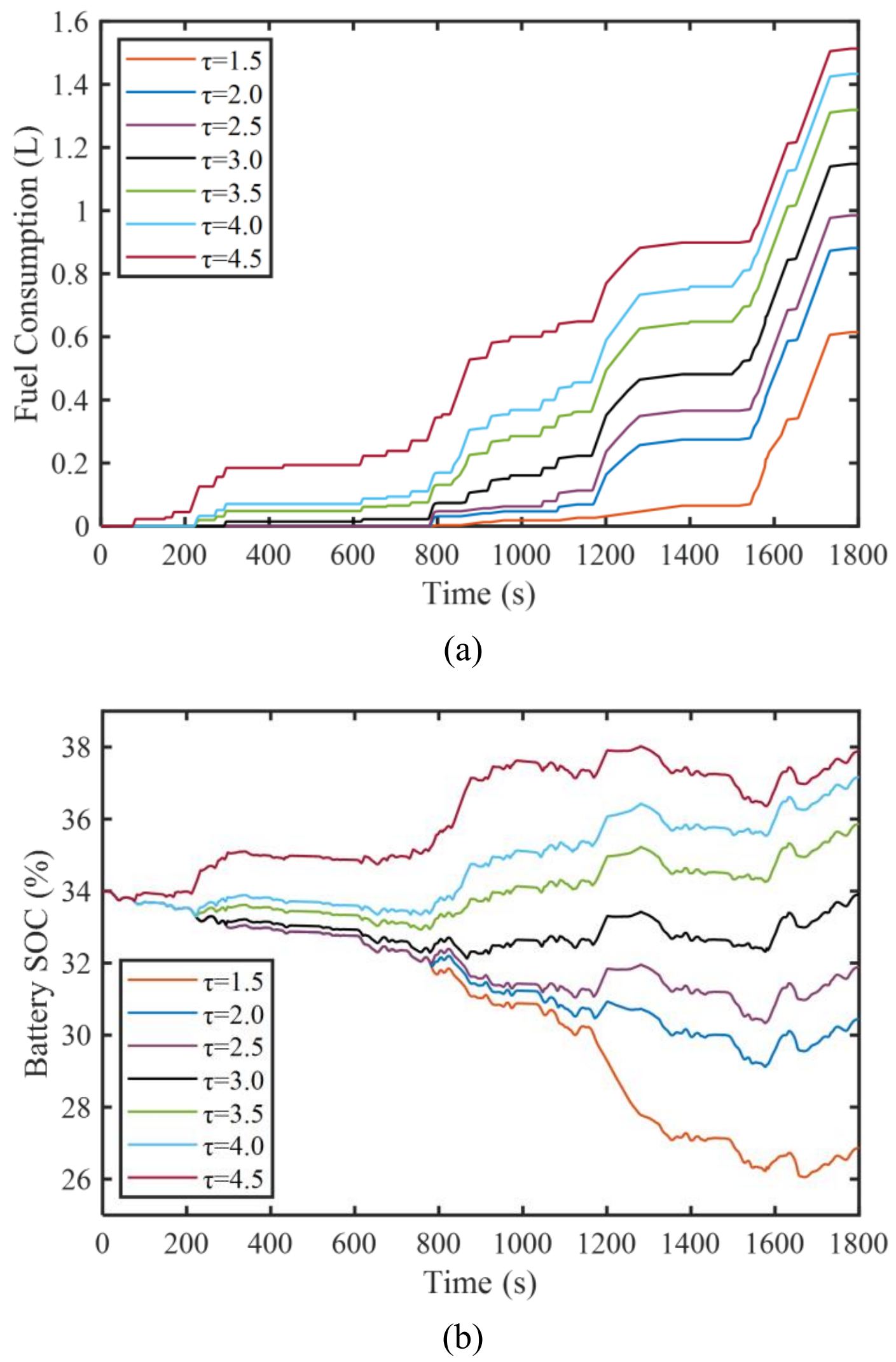

Fig. 17. Comparison of energy management performance in terms of total fuel consumption (a) and battery SOC (b) with different reward function settings.

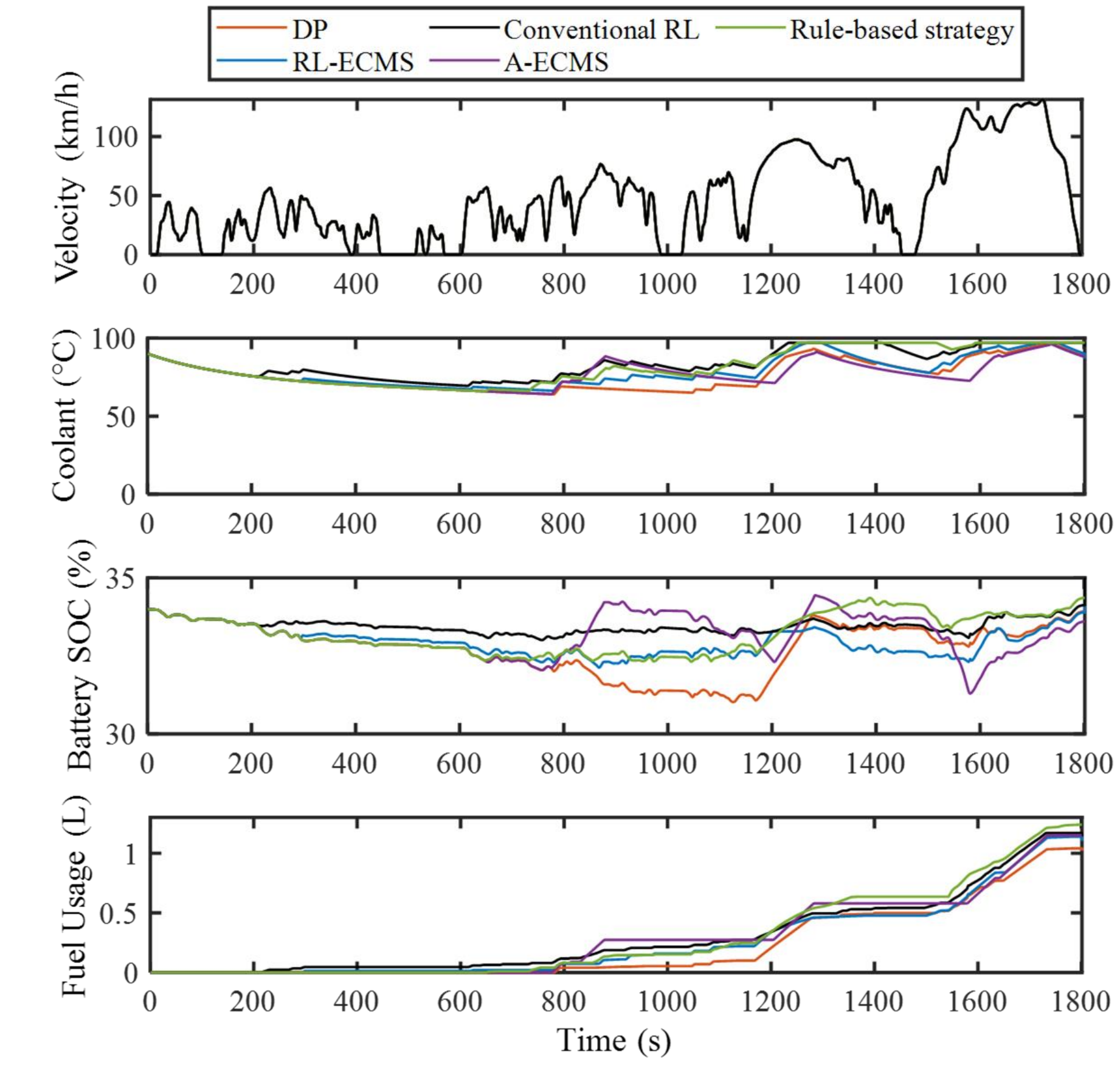

(a)

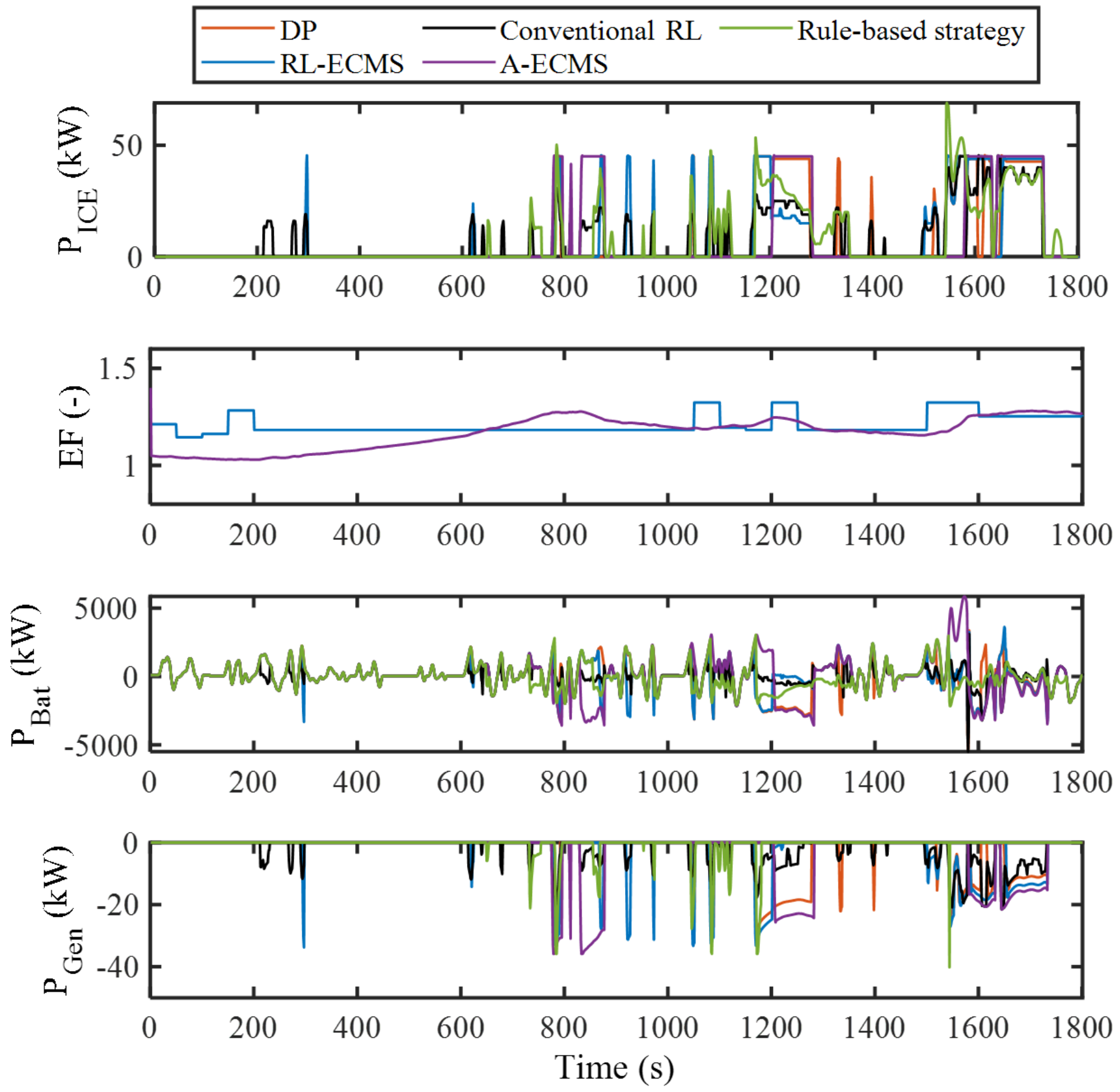

(b)

Fig. 18. Comparisons of PHEV operation profile under the control of different EMS strategies in terms of velocity, coolant temperature, SOC and fuel usage (a) as well as equivalent factor and component output power (b).

**Table 5**
Performance of different EMS control strategies under test driving cycle.

| Performance indicator | DP | RL-ECMS | RL | A-ECMS | RB |
|---|---|---|---|---|---|
| Battery final SOC/ % | 33.99 | 33.96 | 34.20 | 33.67 | 34.42 |
| Fuel consumption/ L | 1.06 | 1.14 | 1.16 | 1.15 | 1.22 |
| Fuel economy/ L/100 km | 4.56 | 4.90 | 4.99 | 4.95 | 5.25 |
| Corrected fuel economy/ L/100 km | 4.58 | 4.93 | 4.96 | 5.03 | 5.17 |
| Engine start-stop counts/ - | 7 | 12 | 23 | 6 | 14 |
| Fuel savings/ % | 13.19 | 5.40 | 4.63 | 3.06 | - |

In Fig. 19, RL-ECMS shows relatively stable characteristics in terms of equivalence factor adjustment. In terms of powertrain component operation, compared to A-ECMS, the absolute values and fluctuations of engine power, battery power, and generator power are larger in A-ECMS due to its PI-based equivalence factor control approach, which struggles to adapt to varying power demand, showing a slower adjustment process. In contrast, RL-ECMS achieves more optimized energy distribution through flexible, rapid adjustment of the equivalence factor. Fig. 20 displays the engine operating profiles of the five energy management strategies under WLTC, showing that the RL-ECMS strategy's engine working points concentrate in areas with higher efficiency.

**Table 6**
The disturbance impacts on RL-ECMS and conventional RL.

| Control strategy | Disturbance / % | ICE power fluctuation / % | Final SOC / % | Fuel consumption / L | Change of fuel consumption / % |
|---|---|---|---|---|---|
| RL-ECMS | 5.0 | 1.07 | 34.67 | 1.17 | 0.43 |
| | 10.0 | 3.23 | 34.98 | 1.19 | 2.15 |
| | 15.0 | 8.76 | 35.13 | 1.21 | 3.86 |
| | 20.0 | 10.02 | 35.30 | 1.22 | 4.72 |
| Conventional RL | 5.0 | 1.12 | 34.06 | 1.18 | 0.61 |
| | 10.0 | 8.70 | 34.03 | 1.20 | 1.12 |
| | 15.0 | 16.67 | 33.89 | 1.14 | 4.04 |
| | 20.0 | 21.75 | 33.81 | 1.11 | 6.62 |

The disturbance rejection capability is further verified as shown in Fig. 21, which demonstrates the performance of the RL-ECMS and the conventional RL strategy when 5%, 10%, 15%, and 20% disturbances are applied to the state inputs. The action outputs and fuel consumption of the two algorithms are depicted in Fig. 21 (a), where the action output for RL-ECMS is the evaluation of the equivalence factor, while the action output for conventional RL strategy is the engine power setpoint.

As the state disturbance increases, the range of action outputs for both strategies also increases with non-linear characteristics. Table 6 summarizes the performance of RL-ECMS and conventional RL strategy under varying levels of disturbance.

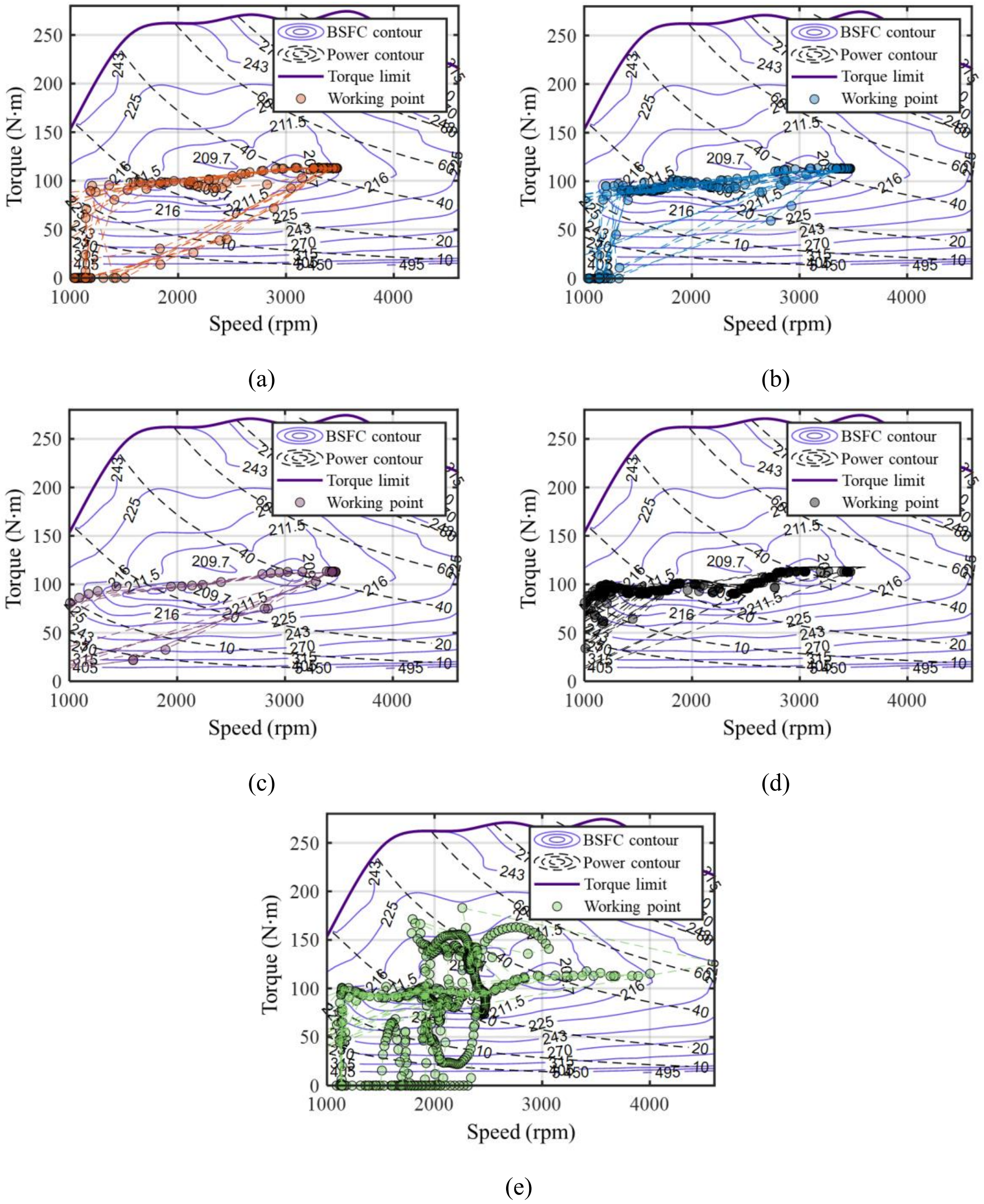

Fig. 20. Engine operation point distribution under the control of different energy management strategies.

Fig. 21 (b) depict the changes in engine power, battery power, and battery SOC for RL-ECMS and conventional RL under different disturbance. With the increase in state disturbance, the SOC deviation for both algorithms tends to rise, but the SOC fluctuation is larger for RL. The average fluctuation of engine power for RL-ECMS increases from 1.07% to 10.02%, the final SOC value gradually increases from 34.67% to 35.30%, fuel consumption rises from 1.17 L to 1.22 L, and the fuel consumption change increases from 0.43% to 4.72%. In contrast, the conventional RL strategy shows greater fluctuations under the same disturbance conditions, with average engine power fluctuation sharply increasing from 1.12% to 21.75%, the final SOC value showing a downward trend from 34.06% to 33.81%, and fuel consumption experiencing an initial increase followed by a decrease, from 1.18 L to 1.11 L, with the fuel consumption change significantly increasing from 0.61% to 6.62%. This demonstrates that the RL-ECMS has better disturbance rejection capability than the conventional RL strategy, showing that RL-ECMS can achieve reliable energy management for PHEVs even under strong state disturbances or uncertainty in environment observation.

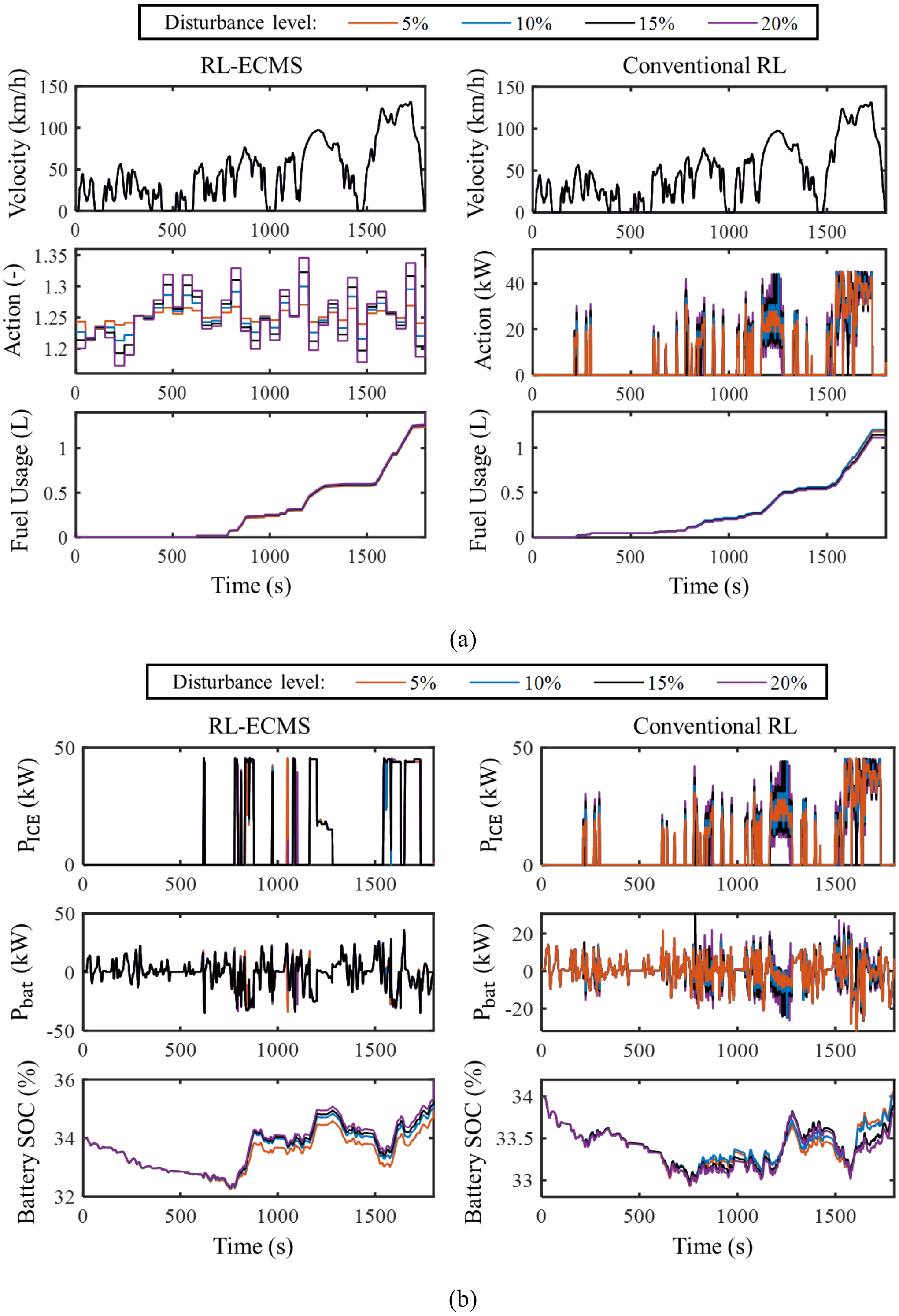

Fig. 21. Performance of RL-ECMS and conventional RL when state disturbance is applied: action output and fuel consumption (a), and component output power and battery SOC (b).

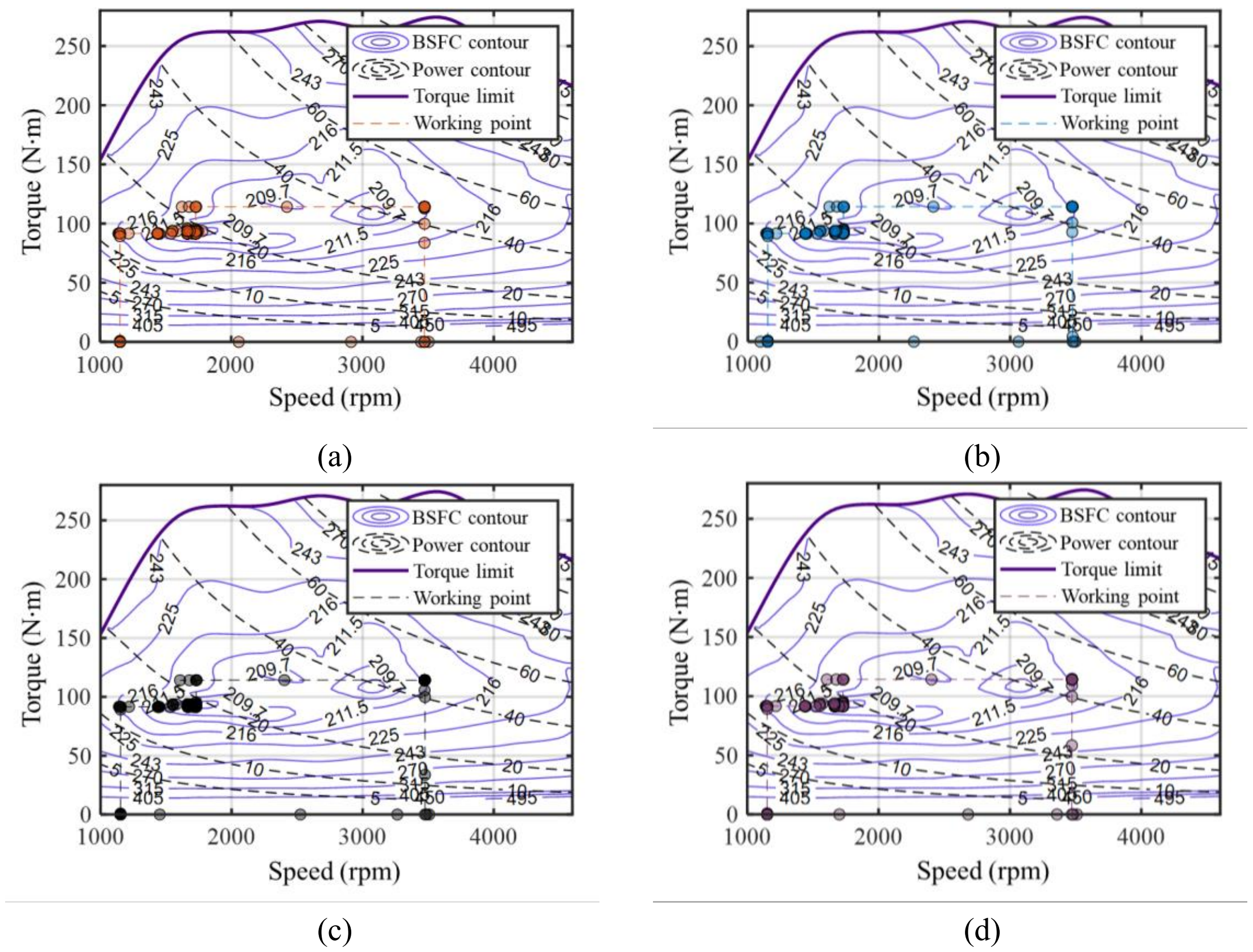

Fig. 22. Engine operating point under the control of RL-ECMS during 1100s to 1300s for DP (a), RL-ECMS, conventional RL (c), A-ECMS (d).

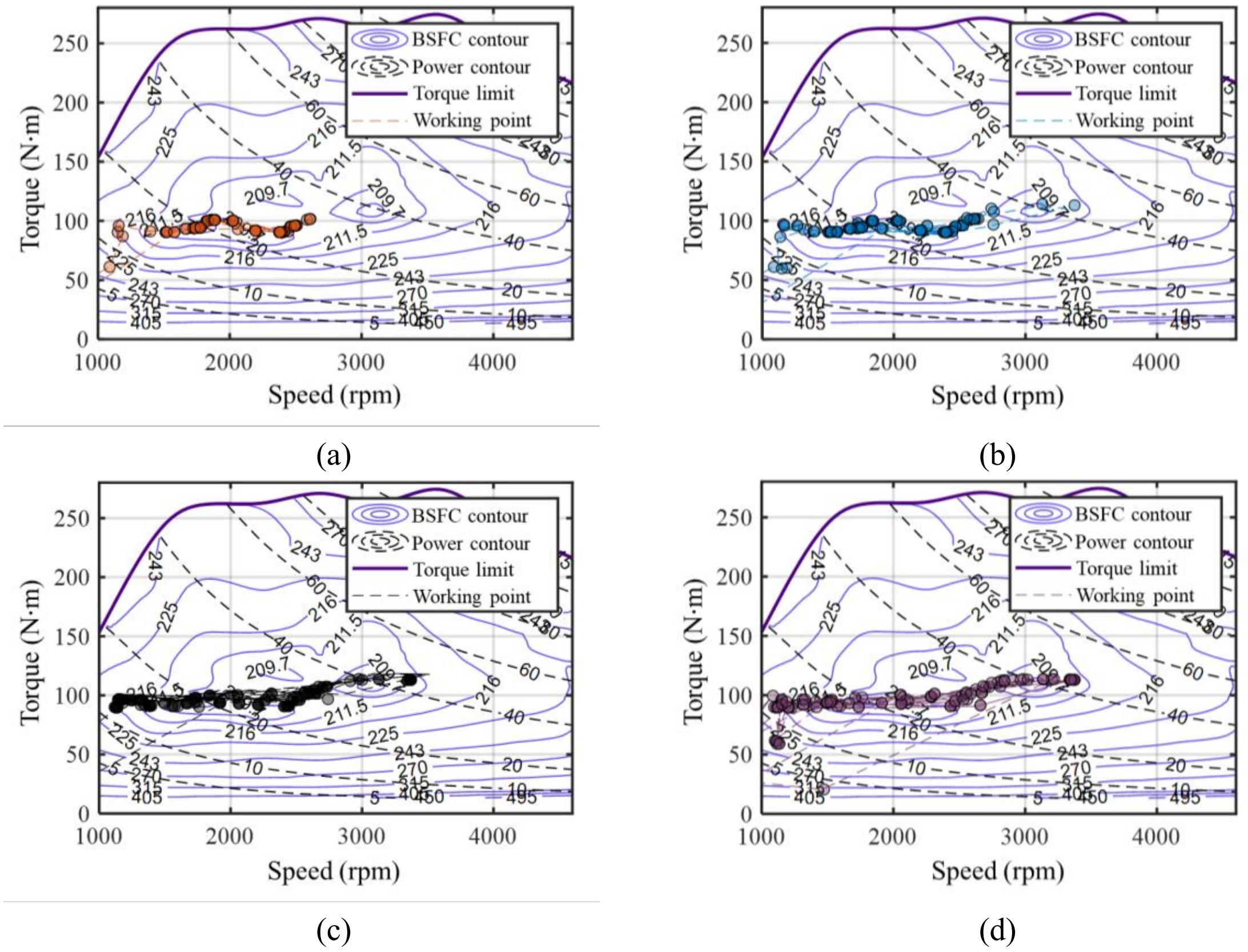

Fig. 23. Engine operating point under the control of the conventional RL strategy during 1100s to 1300s for DP (a), RL-ECMS, conventional RL (c), A-ECMS (d).

Fig. 22 and Fig. 23 compares the engine operation profile under 5%, 10%, 15%, and 20% state disturbances for RL-ECMS and conventional RL between 1100 seconds and 1300 seconds. It is evident that even as the disturbance level increases, the engine working points within the high-efficiency area controlled by the RL-ECMS energy management strategy remain relatively stable without significant dispersion or deviation. This further verifies the effectiveness of the RL-ECMS strategy in resisting environmental state disturbances or observation errors, maintaining stable and safe engine operation. This disturbance resistance is a necessary guarantee for the application of reinforcement learning in energy management systems, ensuring system reliability and stability under complex operating conditions.

## 6. Conclusions

This work demonstrates the industrial application-oriented prototype development of reinforcement learning-based energy management, enhancing RL applicability through model accuracy and reliable control framework. It begins with the high-fidelity PHEV modeling based on physical and data-driven approaches, then proposes a reliable control framework using horizon-extended reinforcement learning merged with ECMS framework to enhance practical applicability. The performance of the proposed methods have been validated by comprehensive experiments, and the conclusions drawn from the research are as follows:

- This study combines reinforcement learning technology with the ECMS real-time control framework, proposing the RL-ECMS energy management control architecture. The key contribution is to improve the flawed method of equivalent factor evaluation based on instantaneous driving cycle and powertrain states found in existing research, drawing the conclusion that long-time scale driving cycle states are needed by RL for correct equivalent factor evaluation.
- Hardware-in-the-loop tests demonstrate that the proposed control method outperforms the conventional A-ECMS and rule-based strategies in fuel economy by more than 1.99% and 4.64%. Moreover, RL-ECMS achieves similar optimality compared to conventional RL-based energy management systems that directly control powertrain components. The fuel consumptions of RL-ECMS and the conventional RL strategy are 4.93 L/100 km and 4.96 L/100 km, respectively, under WLTC driving cycle.
- When state disturbances are applied to the reinforcement learning agent, with increasing disturbance amplitude, the fuel consumption change rate for RL-ECMS reaches a maximum of 4.72%, and the average fluctuation of engine power peaks at 10.02%. In contrast, under the same disturbances, the traditional RL strategy's fuel consumption change rate peaks at 6.62%, and average engine power fluctuation reaches 21.75%, which shows that RL-ECMS has better disturbance resistance capabilities than conventional RL strategy.

## Acknowledgments

This work was supported by the National Natural Science Foundation of China (Grant No. T2241003), the State Key Laboratory of Intelligent Green Vehicle and Mobility (Grant No. ZZ2023-041) and the Dongfeng Motor Corporation Ltd., China (Grant No. CGSQ2022111518).